# Parity and time-reversal elucidate both decision-making in empirical models and attractor scaling in critical Boolean networks

Jordan C. Rozum[1*], Jorge Gómez Tejeda Zañudo[2,3], Xiao Gan[4,5], Dávid Deritei[6], Réka Albert[1,7*]

[1] Department of Physics, The Pennsylvania State University, University Park, PA 16802, USA
[2] Eli and Edythe L. Broad Institute of MIT and Harvard, Cambridge, MA, 02142, USA
[3] Department of Medical Oncology, Dana-Farber Cancer Institute, Harvard Medical School, Boston, MA, 02115, USA
[4] Network Science Institute and Department of Physics, Northeastern University, Boston, MA 02115, USA
[5] Channing Division of Network Medicine, Department of Medicine, Brigham and Women's Hospital, Harvard Medical School, Boston, MA 02115, USA
[6] Department of Molecular Biology, Semmelweis University, Budapest, Hungary
[7] Department of Biology, The Pennsylvania State University, University Park, PA 16802, USA
[*] Corresponding authors: Email: jcr52@psu.edu (JR), rza1@psu.edu (RA)

Summary: Fast analysis of attractor commitment in networks of stochastic switches answers a 50-year-old attractor scaling question.

## Abstract

We present new applications of parity inversion and time-reversal to the emergence of complex behavior from simple dynamical rules in stochastic discrete models. Our parity-based encoding of causal relationships and time-reversal construction efficiently reveal discrete analogs of stable and unstable manifolds. We demonstrate their predictive power by studying decision-making in systems biology and statistical physics models. These applications underpin a novel attractor identification algorithm implemented for Boolean networks under stochastic dynamics. Its speed enables resolving a longstanding open question of how attractor count in critical random Boolean networks scales with network size, and whether the scaling matches biological observations. Via 80-fold improvement in probed network size (N=16,384), we find the surprisingly low scaling exponent of 0.12±0.05 -- approximately one tenth the analytical upper bound. We demonstrate a general principle: a system's relationship to its time-reversal and state-space inversion constrains its repertoire of emergent behaviors.

## Introduction

Many complex systems in the natural, social or technological realm exhibit emergent behavior, i.e., collective dynamics arising from the interaction of entities governed by simple rules (*1–4*). Examples include phase transitions (*5,6*), flocking (*7*), consensus formation (*8*), and spontaneous synchronization of oscillators (*9*). Modeling frameworks that are frequently used to study the collective behavior of individuals include nonlinear dynamics (*10,11*), agent-based models (*12*), cellular automata (*13*), and network models (*14*, *15*). Boolean models sit at the intersection of these approaches (e.g., *16–18*). They assign a time-varying binary variable to each system entity, represented as a node in a network of interactions. They exhibit diverse long-term dynamics (attractors) that represent collective behavior (for example, consensus of individuals) and they describe the evolution toward an attractor (for example, consensus formation from an initially disordered state).

Boolean modeling of electronic circuits is well-known (*19*), but many other familiar models can also be viewed as

Boolean models. The quenched (zero-temperature) Glauber model, a dynamic variant of the Ising model, considers the dynamics of each atom's two possible spin orientations under the influence of its neighbors (*20*, *21*). Another example category includes models of spreading binary opinions through social networks (reviewed in *8*). The McCulloch & Pitts neural network model introduces a propositional logic of all-or-none neuronal activation (*22*); the Hopfield model also considers two activities for each neuron and assumes a complete network of interactions (*23*).

Boolean models are well suited to elucidate system-level decision-making, i.e., robust commitment towards one of the dynamical attractors in a multi-stable system. This has made their use especially wide-spread in biology. They were introduced by Stuart Kauffman (*24*) and René Thomas (*25*) as prototypical models for gene regulatory networks that underlie cell fate decisions (such as those that happen during cell differentiation). A large body of research has since shown that the attractors of Boolean models correspond to cell fates or stable patterns of cell activity (such as the cell cycle). Boolean models integrate and encode current knowledge of a biological process, fill any gaps of knowledge with hypothesized interactions, and predict the behavior of the system under loss-of-function, constitutive activation, or external control of system entities (*25–27*). They are frequently used to study cell differentiation processes such as T cell specialization (*28*, *29*), developmental processes such as patterning during embryogenesis in *Drosophila melanogaster* (*30*, *31*), and cancer (e.g., metastatic reprogramming (*32–34*), prediction of targeted therapies (*35*, *36*)). Model predictions in a variety of systems were verified experimentally (*34*, *37–40*).

Alongside models of specific systems, analysis of the expected collective behaviors exhibited by generic Boolean models has also proven insightful. Ensembles of Boolean models (Random Boolean Networks) have been studied for decades (reviewed in *41–43*). These ensembles exhibit an order to chaos transition as dynamical and topological parameters are tuned. In the intermediate (critical) regime the ensembles exhibit features of biological cells, including stability against perturbations and plausible scaling laws for the number and size of attractors with the system size (see Text S1 for details). Here we answer a long-standing question about the scaling law in the presence of timing stochasticity (i.e., when the update order and timing of variables is stochastic).

Despite Boolean models' discrete nature and apparent simplicity, it is nontrivial to connect dynamical properties of decision-making to the underlying interaction network. Brute-force exploration of their state spaces is not generally feasible. A typical Boolean model of a biological process with a few dozen variables has tens of billions of states. Genome-scale models can have thousands of variables, resulting in many more states ($\sim 10^{300}$ $to$ $\sim 10^{9,000}$) than Planck volumes in the observable universe ($\sim 10^{185}$). This challenge has motivated decades of research analyzing discrete dynamics without exhaustive state-space searches (*44*), for example by analyzing how feedback loops in the interaction network constrain dynamics (*45*, *46*). While the body of research regarding how network structure constrains dynamics has proven invaluable, it is important to note that multiple Boolean systems are compatible with each interaction network. Ambiguity can be eliminated by defining a network whose graph structure unambiguously represents the update functions that govern the time evolution of each variable. One such network representation, the *expanded network* (also called the logical or prime-implicant hypergraph) (*31*, *47*, *48*), defines two *virtual nodes* for each entity, denoting the two possible values of its binary variable. Connections among virtual nodes encode the update functions. The structure of this auxiliary network tightly constrains the attractors of a system (*33*, *49*, *50*). The expanded network can be used to identify control strategies that drive the system to a desired attractor (*51*, *52*). The most parsimonious of these control strategies involve determining the so-called driver node(s) of self-sustaining circuits (*53*, *54*). In this way, fixing the state of a few driver nodes ensures the system's convergence into a target space from any initial condition. This type of analysis has been used, for example, to pinpoint key proteins in pathological cell processes (*51*, *53*, *55*). Although initially developed for Boolean models, the expanded network has been generalized to multi-level discrete and ODE analogs (*55*, *56*).

We go beyond the previous use of the expanded network and characterize each virtual node with a binary activity and impose a parity structure that facilitates the proof of new theorems about the effects of perturbations on system trajectories. We call this extended version of the expanded network the *parity-expanded network*. With these additions, the parity-expanded network is a complete representation of the Boolean dynamical system, and is a dynamical system in its own right. In addition, we describe the time-reversal of stochastically asynchronous Boolean systems and use it to identify subsets of the state space that cannot be reached from states outside the subset. Using parity and time-reversal transformations in tandem, we developed a new algorithm to efficiently identify all attractors of large-scale Boolean systems. We apply the algorithm to answer the long-standing question of how quickly the number of attractors in asynchronous Random Boolean Networks increases with network size.

## Materials and Methods

We begin by recalling relevant concepts from Boolean modeling and defining the notation we use throughout. Constructing a Boolean model usually starts with the synthesis of the modeled system's interaction graph. An interaction graph is a signed directed graph whose nodes are the $N$ entities of a system and whose edges represent positive (activating) or negative (inhibiting) influence. Each entity $i$ is characterized by a variable $X_i$ that can take one of two values: 1 ("active") or 0 ("inactive"). Each $X_i$ updates its value according to the output of an *update function* $f_i: \{0,1\}^N \to \{0,1\}$ which maps each system state $\boldsymbol{X} = (X_{i_0}, \dots, X_{i_{N-1}})$ to either 1 or 0. Any Boolean function can be expressed algebraically using logical operators (e.g., "$not$", "$or$", "$and$" with symbolic representations ¬, ∨, and ∧, respectively). There are several schemes for determining the timing of variable updates. We use the *stochastic asynchronous update* scheme, in which at each step a single variable is randomly chosen to update its value (each variable must have a non-zero update probability; these are often chosen to be uniform). Compared with other updating schemes, this scheme removes spurious oscillations that arise from unrealistic perfect synchrony, but otherwise preserves long term behaviors (*57–59*). Once the update functions are determined and an update scheme is selected, the Boolean system is fully specified. Throughout this work we use the stochastic asynchronous update scheme, and as such, the term "system" is to mean a set of $N$ update functions $f_i$ together with the implicit stochastic asynchronous update scheme. Each Boolean system induces a *state transition graph* (STG) with $2^N$ nodes that represent all possible system states and with directed edges from one node (system state) to another when the parent state can be updated in one time-step to attain the child state. Under stochastic asynchronous update, each node has between 0 and $N$ outgoing edges. The *attractors* of a Boolean system are the terminal strongly connected components of the STG (i.e., they have no edges that exit the component). They are divided into two types: point attractors (also called fixed points or steady states), which contain only one state, and complex attractors, which contain more than one state. Importantly, the topology of the STG is not affected by biasing some nodes to update more frequently than others. Therefore, the attractor repertoire does not depend on the precise probabilities that individual nodes are selected for update in the stochastic asynchronous update, as long as the probabilities remain nonzero.

### A new framework: The expanded network through the lens of parity

The *expanded network* (also called the logical or prime-implicant hypergraph) (*31*, *47*, *48*) was introduced as an auxiliary network constructed from the Boolean update functions. The expanded network nodes represent Boolean literals (e.g., $X_i$ and $\neg X_i$) and its hyperedges (generalized edges that connect sets of nodes) represent prime implicants (irreducible sets of regulator states that result in $f_i(\boldsymbol{X}) = 1$) of the update functions. Here we introduce a new definition of the expanded network that uses parity-related concepts and highlights its role as an invariant of the parity transformation.

The *parity transformation* acts on a Boolean system by the change of variables $X_i \mapsto \neg X_i$ for all variables $X_i$, mapping the original system with update functions $f_i$ to the system governed by update functions $\neg f_i$. This mapping induces further transformations on any structure derived from the update functions, and so, in a slight abuse of notation, we say the parity transformation acts on all these structures. For example, the parity transformation relabels the nodes of the state transition graph so that all 1s become 0s and vice versa (see Figure 1). Viewing these node labels as spatial coordinates (so states lie on the vertices of a unit hypercube), the parity transformation is the spatial inversion of this hypercube through its center (see Figure 1).

Parity allows for a succinct definition and extension of the expanded network that builds upon the eponymous structure defined in (*48*). Here, we give an abbreviated discussion of this object and leave formal details to Text S2. A parity-expanded network $G$ is a dynamically endowed hypergraph. Each node $I$ of the parity-expanded network, called a *virtual node*, is an ordered pair $I = (n(I), s(I))$ consisting of a system entity, in this context denoted $n(I)$, and a value $s(I)$, which is either the constant 1 or the constant 0. There are two virtual nodes associated with each system entity $i$, namely $(i, 1)$ and $(i, 0)$; we call this pair of virtual nodes *contradictory*. A set of virtual nodes that does not contain any contradictory pair is called *consistent*. Each virtual node $I$ is endowed with a Boolean variable $\sigma_I$, whose time evolution is governed by an update function $F_I$; this defines a $2N$-dimensional dynamics which can be restricted to an $N$-dimensional subset such that $F_{(i,0)}(X) = F_{(i,0)}((\neg X, X)) = \neg f_i(X)$, and $F_{(i,1)}(X) = F_{(i,1)}(\neg X, X) = f_i(X)$. In the context of this restriction, we may think of $\sigma_I$ as the indicator function for the subspace defined by $I$, i.e., we define $\sigma_{(i,0)}(X) = \neg X_i$, and $\sigma_{(i,1)}(X) = X_i$. We view the expanded network as having two layers: an "original update" layer and a "parity update" layer. This can be seen in Figure 1, where the parity-expanded network nodes are partitioned into these two layers. In contrast with earlier versions of the expanded network and the logic hypergraph, the parity-expanded network is a dynamical system in its own right; its nodes are characterized with activity variables and a stochastic time evolution function $\boldsymbol{F}: \{0,1\}^{2N} \to \{0,1\}^{2N}$, restricted to the $N$ degrees of freedom of the underlying Boolean system. See Text S2 for an example of $\boldsymbol{F}$ written explicitly as a function of $2N$ variables for the network of Figure 1. A main advantage of this notation is that it allows us to treat the negated and non-negated versions of variables and functions simultaneously. In addition, the explicit endowment of the parity-expanded network with dynamical properties avoids awkward constructions along the lines of "the Boolean network corresponding to the parity-expanded network" in several places; rather, we can speak more simply of just "the parity-expanded network".

The dynamics of these activity variables are encoded in the connectivity of the parity-expanded network. A *hyperedge* connects a set of parent virtual nodes $S = \{I_0, I_1, \ldots, I_k\}$ to a target virtual node $J$ if $\wedge_{I \in S} \sigma_I$ is a prime implicant of the update function for $\sigma_J$, i.e., of $F_J$. Pictorially, we represent hyperedges with more than one parent using intermediary "composite nodes", which correspond to "$and$" gates. For an example of a Boolean system and its parity-expanded network, see Figure 1B, C. Hyperedges between and within parity layers encode important features of the dynamics. For example, negative influence manifests as inter-layer hyperedges. Thus, it follows from Theorem 19 of (*46*) that if a Boolean system's interaction graph lacks negative feedback loops and has no paths of opposite sign between any two nodes, then there is a change of variables that disconnects the parity layers from one another.

Arguably the most dynamically important of the parity-expanded network's topological structures are its *stable motifs* (*48*) and *stable modules* (*56*), which correspond to specific states of generalized positive feedback loops in the interaction graph (though we will define them on the parity-expanded network). These determine *trap spaces* in the dynamics, which are regions of the state-space characterized by a set of fixed variable values that, once attained by a trajectory, confine the trajectory to that region for all subsequent time-steps. These generalize the notion of point attractors in that only a subset of the system's variables is fixed in a trap space. Many of our formal results rely on recasting these structures in the parity view of the parity-expanded network as follows. A stable module $M$ is a non-empty sourceless sub-hypergraph of the parity-expanded network such that $M$ does not overlap (does not share virtual nodes) with its image under parity. A stable motif is a stable module that does not contain any smaller stable module; note that this implies that a stable motif is strongly connected. Because stable modules (and thus also stable motifs) are sourceless, every virtual node in a stable module $M$ can be maintained in its active state by other virtual nodes in $M$, meaning that the activity of $M$ is self-sustaining. Once $M$ is activated, it cannot be inactivated except via direct override of its virtual node activities by direct external controls (as opposed to inactivation via the effects of upstream pathways). Thus, $M$ describes a control-robust trap space in which the values of certain variables are stationary (*47*, *51*). The trap spaces corresponding to the activity of stable modules are exactly those considered by (*49*), with larger trap spaces (more states) corresponding to smaller stable modules (fewer constrained variables) and vice-versa. See Figure 1 for an example of the parity-expanded network and stable motifs.

We leverage the parity properties of the parity-expanded network to prove new results about driver node sets (*54*) and their relation to attractors. Formal statements and proofs are given in Text S3. Recall that a set of virtual nodes is called consistent if it is disjoint from its image under parity (i.e., it contains no pair of nodes of the form $(i, 1)$ and $(i, 0)$). We say that a consistent set of virtual nodes $S$ *drives* a virtual node $I$ if $I$ is consistent with $S$ and if $F_I(X) = 1$ for every attractor state $X$ of the dynamics obtained by restriction to the states in which $S$ is active. As a particular example, the vertex set of any stable motif drives itself (is self-sustaining). The set of all virtual nodes driven by $S$ is called the *domain of influence* (DOI) of $S$, written $DOI(S)$. It follows from this definition that (with probability 1) trajectories with $S$ initially active eventually either inactivate $S$ or activate all of $DOI(S)$. It is often useful to study the subset $DOI(S) - S$ of $DOI(S)$, which consists of all virtual nodes $(i, s) \notin S$ for which $X_i = s$ is fixed in all attractors of the dynamics restricted to the subspace defined by $S$. Similarly, $DOI(S)$ in its entirety is the union of $DOI(S) - S$ and the members of $S$ driven by $DOI(S) - S$. We say that $S$ is *self-negating* if there exists a subset $T$ of $DOI(S) - S$ such that $DOI(T)$ intersects the image of $S$ under parity (i.e., $T$ drives a node that contradicts $S$). In such cases, $S$ cannot be active in every state of any attractor. This definition of the domain of influence is closely related to concepts presented in (*54*).

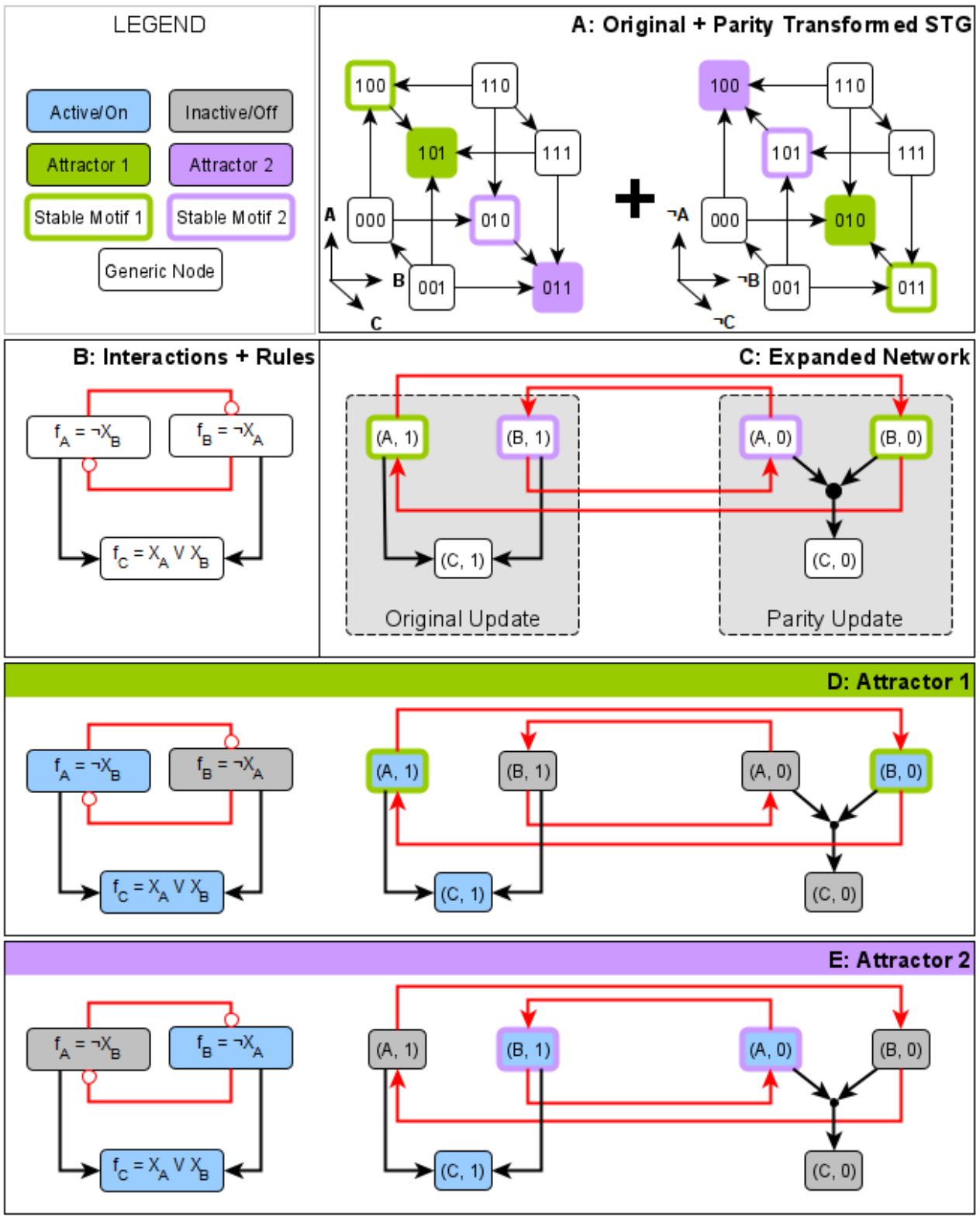

Figure 1: The relationship between the parity-expanded network and parity transformation illustrated on a three-variable Boolean system. The state transition graphs are shown in panel A; each system state is represented as the triple $X_A X_B X_C$. Each state with fewer than three outgoing edges (state transitions) also has a self-loop, which is omitted for visual clarity. The interaction network and update functions are indicated in panel B and the parity-expanded network in panel C. The system has two attractors, shown in panels D and E, with blue nodes active and grey nodes inactive. States in the state transition graphs of panel A are arranged so that the individual variables define coordinate axes. In this arrangement, the states form the corners of a cube. The parity transformation reflects each attractor through the center of this cube. The parity-expanded network in panel C has two parts, or layers: regular virtual nodes (A,1), (B,1) and (C,1), whose activity updates according to the usual update functions, and negated virtual nodes (A,0), (B,0) and (C,0) that use the parity transformed update functions. The filled black circle represents the hyperedge from the set {(A,0), (B,0)} to (C,0) and indicates the "$and$" operation in the update function of the virtual node (C,0). Positive regulation (black arrows) stays within a layer of the parity-expanded network, while negative regulation (red arrows) crosses between layers. Virtual nodes (A,1) and (B,0) form a stable motif (green outline), as do the virtual nodes (A,0) and (B,1) (purple outline). In any state of the system, half of the parity-expanded network is active. The stable motif $(A, 1), (B, 0)$ describes the trap space containing Attractor 1 and does not overlap with its parity transformed set $(A, 0), (B, 1)$, which corresponds to the trap space containing Attractor 2 (panel E).

Calculating $DOI(S)$ can be difficult in general, so we focus instead on a commonly used and easily calculated subset of the driving relation: the logical domain of influence (LDOI). A set $S$ of virtual nodes *logically drives* a virtual node $I$ (which may or may not be in $S$) if there is a non-trivial multipath in the parity-expanded network from a subset of $S$ to $I$ with all virtual nodes in the path consistent with $S$ (this requires that $I$ is consistent with $S$). Here, a (non-trivial) multipath from a set $S$ to a node $I$ is a (non-empty) finite sequence of hyperedges $\{(h_{parents,i}, h_{children,i}): i = 0,1, \ldots, n\}$ such that i) $h_{parents,0} \subseteq S = S_0$, ii) $h_{parents,i} \subseteq S_{i-1} \subseteq h_{children_{i-1}}$, and iii) $I \in h_{children,n}$. Note that for $I \in S$, a consistent non-trivial multipath exists from $S$ to $I$ (i.e., $I \in LDOI(S)$) if and only if upon restriction to the subspace defined by $S$ and percolation of constant values, $I$ becomes active. The set of all $I$ logically driven by $S$ is called the *logical domain of influence* of $S$, written $LDOI(S)$ (or $LDOI(i, s)$ when $S = \{(i, s)\}$ is of size one). Intuitively, $LDOI(S)$ corresponds to the variable values that become fixed after percolating $S$ through the update functions and simplifying algebraically. As demonstrated in (*54*), if $S$ is a subset of $LDOI(S)$, then $LDOI(S)$ contains the virtual nodes of a stable motif. If $LDOI(S)$ contains a stable motif $M$, we say that $S$ (logically) drives $M$.

Our main result in this section (Theorem 1, Text S3) states that if an attractor contains a state for which $S$ is active, it must also contain a state for which $DOI(S)$ is active. This result illustrates how the domain of influence of a set of virtual nodes can constrain which states must coexist within an attractor. Such considerations are important in constructing Boolean models in which certain system configurations should correspond to different long-term qualitative system behaviors (e.g., phenotypes). Two corollaries of this result allow one to study the conditions under which an attractor avoids activating a particular set of stable motifs. This problem is of interest, for example, in biology, where one or more stable motifs may correspond to a diseased state of the system; the goal in this case is to identify drug targets that avoid stabilizing the diseased state. We will also make use of these results in later sections to enumerate a system's attractors.

The first of the two corollaries (Corollary 1, Text S3) can be viewed as a compatibility condition for a stable motif's activity and the activity of a virtual node that drives it. In the context of system control, it states that if fixing $X_i$ is sufficient to eventually activate a stable motif $M$, then the oscillation of $X_i$ is also sufficient to activate $M$. It can also be viewed as a consistency condition for attractors: if $X_i = s$ leads to activation of $M$, then we cannot have $X_i = s$ in an attractor (even transiently) in which $M$ is inactive. This result provides a powerful way to identify circumstances in which no stable motif activates: in such cases, all stable motif drivers must be permanently inactive. Specifically, we collect all single-node drivers of all stable motifs into a set $\Delta$, and test whether or not $\neg \Delta = (i, \neg s): (i, s) \in \Delta$ is self-negating. If it is self-negating, then it follows that at least one element of $\neg \Delta$ is not permanently active in each attractor, and thus (by Corollary 1) this element must eventually stabilize at least one stable motif. The formal statement of this result is given as Corollary 2 in Text S3.

## Time-reversal of Boolean systems and stable motifs of the time-reversed system as unstable "Gardens of Eden"

A second set of foundational results in this work is the construction of the time-reversal of an asynchronous-update Boolean system -- which exists despite the system's inherent stochasticity. We use the time-reversal transformation to help identify discrete analogs of unstable manifolds in the state space. The *time-reversal* (TR) of a Boolean system governed by update functions $f_i$ is called the time-reversed system and is governed by the update functions $f_i^-$, where $f_i^- = \neg f_i(X_i = \neg X_i)$ (i.e., in a state $X$, the value of $f_i$ is obtained by negating the value of $X_i$, evaluating $f_i$, and taking the negation of that output). Like the parity transformation, we formally define the time-reversal as a transformation of the Boolean system given by its effect on the update functions. It similarly induces transformations of structures that are derived from these functions. For example, the time-reversal of a parity-expanded network $G$ for update functions $f_i$ can be obtained as the parity-expanded network for the update functions $f_i^-$, and is denoted $TR(G)$. Similarly, the state transition graph of a system is related to its time-reversed counterpart by reversing the direction of all edges (see Figure 2A). Thus, one may follow the evolution of the time-reversed system on the STG by following edges in the reverse direction. Equivalently, and in analogy to concepts in solid-state physics, one may imagine that all but one of the nodes in the STG are occupied by walkers who may not share an STG node. If the walkers randomly follow the edges of the STG, then the position of the unoccupied "hole" evolves according to the possible trajectories of the time-reversed system.

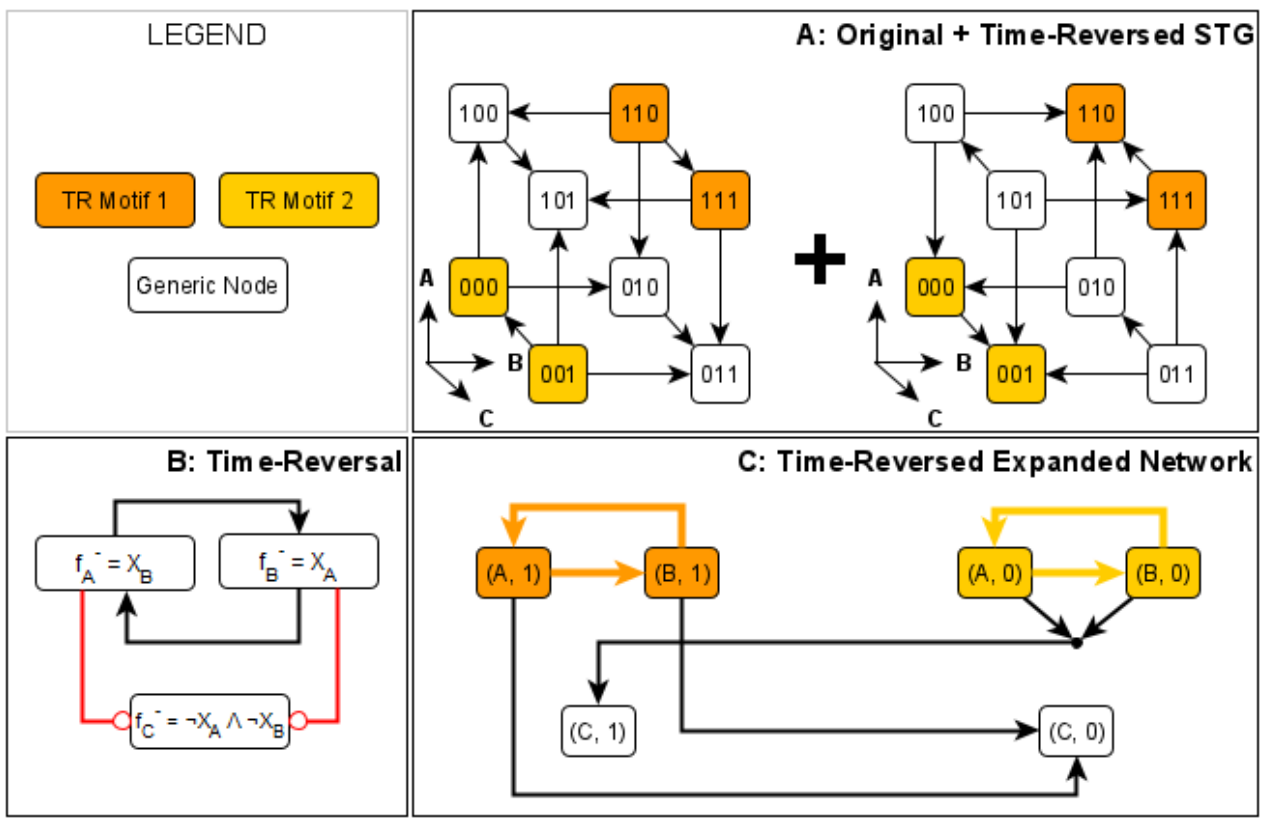

Figure 2: Illustration of the time-reversal transformation on the example from Figure 1. Panel A shows the original system's STG alongside the STG of its time-reversal. The time-reversal has the effect of reversing the direction of each edge in the STG. The interaction network and update functions of the time-reversed system are depicted in panel B alongside its parity-expanded network in panel C. Red interaction network edges indicate inhibition, while black edges indicate activation. The virtual nodes and hyperedges of the time-reversed system's two stable motifs are highlighted in yellow and orange in the parity-expanded network, and the regions of state space in which they are active are highlighted in the system's STG on the left in panel A. Note that the system states highlighted in yellow (000 and 001) viewed in the original system's STG form a subgraph that has no in-component and the state 001 is a Garden of Eden state. The same is true for the system states highlighted in orange. In this example, the sign of the regulation (inhibition vs activation) is reversed under the time-reversal. This holds in general for all interactions except self-regulation, which does not change sign under time-reversal.

The concept of Garden of Eden states, which are source nodes of the STG (*60*, *61*), can be generalized as subgraphs of the STG that do not have incoming edges; we call these subgraphs Garden of Eden spaces. They are analogous to unstable manifolds of ODE systems in the sense that no trajectory can enter these spaces from the outside. Any trap space of a system, and in particular any of its stable motifs, is a Garden of Eden space in the time-reversed system, and vice versa. For example, the states marked in green in Figure 1A form a trap space of the original system and a Garden of Eden space of its time-reversed counterpart while the states marked in yellow form a trap space of the time-reversed system and a Garden of Eden space of the original system. An important consequence of this time-reversal-based mapping between trap spaces and Garden of Eden spaces is that no attractor of a system can cross the boundary of any of the system's trap spaces or Garden of Eden spaces. This observation is especially helpful in eliminating states from consideration when searching for attractors via direct STG construction, or in reducing the number of relevant initial conditions for study. We leverage these methods to construct an efficient attractor-identification algorithm and explore their utility by way of example in the section "Application to decision-making in empirical biological network models".

## Stable motif succession determines state-space decision-making and attractors

As a Boolean system's state transition graph (STG) has $2^N$ nodes and up to $2^N N$ edges (for stochastic asynchronous update), for systems with many entities (large $N$) it is impractical to use the state transition graph to determine the system's attractor repertoire. Stable motif, or trap space, based attractor identification methods are often more effective (*47*, *48*). In the iterative approach of (*48*), a system's stable motifs are identified, and one is selected to "lock in". The system's update functions are then reduced under the assumption that the system's state is confined to the region described by the stable motif, resulting in a *reduced network*. Rephrased in our framework, each stable motif is selected in turn and the parity-expanded network is simplified under the assumption that the motif's virtual nodes are active, resulting in a reduced Boolean system. We use the notation $Red(G, M)$ for the reduced parity-expanded network that results after restricting the dynamics to the subspace defined by the activity of $M$ and its LDOI. The process is repeated recursively for each reduced parity-expanded network until all possible permutations of stable motif activation are explored. The result is a *succession diagram, $\Sigma$*, which is a directed acyclic graph whose nodes are the unions of the vertex sets of stable motifs used to obtain each reduced system (see Text S2 for a formal definition). For an example of this process and the resulting succession diagram, see Figure 3.

The succession diagram serves as a summary of the decisions in the system dynamics that lead to successively more restrictive nested trap spaces. Each node of the succession diagram corresponds to a region of the state space in which the denoted stable motifs are active. Each branch point in the succession diagram represents potential choices to be made; which choice is ultimately selected by the system depends on various factors including the stochastic update order. It follows from the nestedness of these trap spaces that the system cannot transition between regions that are not connected by a path in the succession diagram, i.e., the succession diagram encompasses the entire repertoire of decisions the system is capable of making. The close relationship between the succession diagram and branch points in the dynamics is illustrated in Figure S1 by constructing the full state transition graph of the example from Figure 3. In the section "Application to decision-making in empirical biological network models" below, we illustrate how the succession diagram can be used to analyze the complex state space decision-making in systems biology models.

One must take special care to consider the possibility of oscillations that avoid activation of stable motifs. For example, consider the system shown in Figure 4

$$f_A(X) = f_B(X) = \neg X_A \wedge \neg X_B \vee X_C; f_C(X) = X_A \wedge X_B.$$

This system's parity-expanded network contains only one stable motif, the hypergraph induced by $(A, 1), (B, 1), (C, 1)$, which corresponds to the system's sole point attractor $X_A = X_B = X_C = 1$. Previous stable motif or trap-space based methods (*48*, *49*) would correctly identify this point attractor by finding the corresponding stable motif. This system, however, contains an additional attractor in stochastic asynchronous update. In the second attractor, $X_C$ remains in the 0 state, while $X_A$ and $X_B$ oscillate; this second attractor is not identified by previous iterative stable motif reduction methods. Other existing methods can identify this attractor via simulation or can detect that at least one attractor lies outside the union of identified trap spaces. For a detailed discussion of the nature of these oscillations and their implications for the completeness of the attractor repertoire identified by iterative stable motif reduction, see (*51*). Such oscillations motivate us to propose more robust automated methods that can identify oscillations that fail to activate stable motifs. These methods make practical what previously was impractical: the identification of the attractor repertoire of ensembles of large Boolean systems. Importantly, our method automatically identifies all complex attractors, including those that were overlooked by previous iterative stable motif reduction methods.

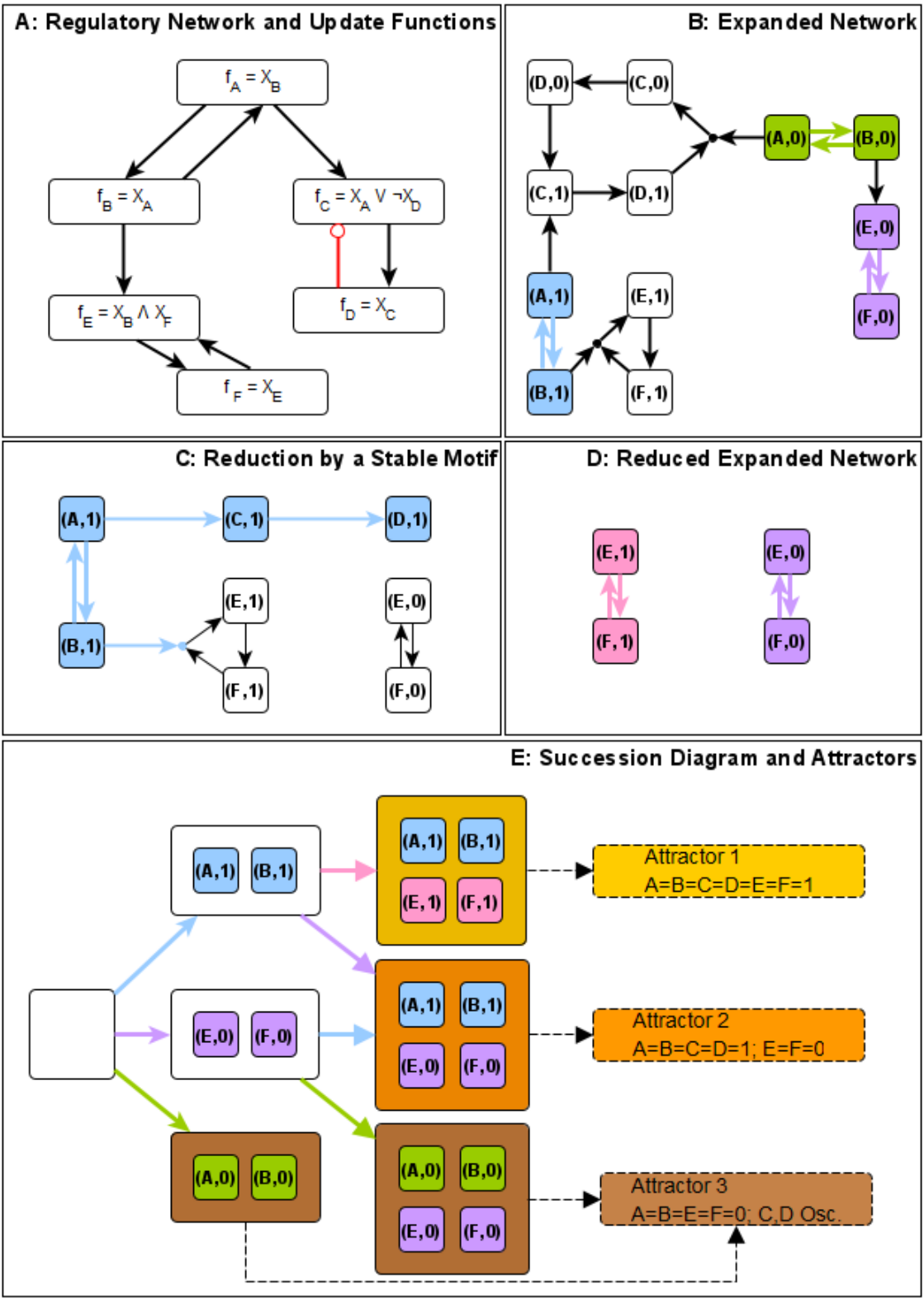

Figure 3: Outline of the iterative stable motif reduction process on a simple example. Panel A depicts a Boolean system's interaction network and update functions. The corresponding parity-expanded network, $G$, is shown in panel B; there are three stable motifs, highlighted in blue, green or purple. In panel C, the stable motif corresponding to $X_A = X_B = 1$ is selected and the effect of maintaining its activity is highlighted in blue; in particular, it leads to $X_C = X_D = 1$. The variables that are unfixed ($X_E$ and $X_F$) then form the bistable system $Red(G, \{(A, 1), (B, 1)\})$ whose parity-expanded network is shown in panel D. The reduced system has two stable motifs, highlighted in pink or purple; the latter was also a stable motif of the original system. All possible sequences of stable motif selection and reduction are summarized in a succession diagram (panel E). Each node of the succession diagram is a set of virtual nodes that contains the stable motifs that were selected for use in the reduction. The colors of the edges in the succession diagram indicate which stable motif is selected to get from one reduction to the next. The process terminates when no stable motifs remain, and the attractors of the maximally reduced systems are identified as attractors of the unreduced system (highlighted in yellow, orange, and brown).

## Overview of the attractor identification algorithm

We follow an iterative stable motif approach to attractor identification in which stable motifs are recursively used to produce reduced Boolean systems (and corresponding reduced parity-expanded networks) until no additional stable motifs remain (see Figure 3). At each stage in the iteration for which a reduced parity-expanded network contains a stable motif, we identify complex attractors that do not activate any additional stable motifs. We call such attractors *motif-avoidant* and call reduced systems with motif-avoidant attractors *terminal*. Though terminality requires analysis of the system's state transition graph in the general case, it is usually possible to significantly reduce the computational burden by application of a necessary condition for terminality that arises from Corollary 2 (Text S3) and properties of the parity-expanded network. We consider the set $\Delta$ of all virtual nodes that individually drive any stable motif to obtain Theorem 3 (Text S3), which can be informally stated as follows: all motif-avoidant attractor states satisfy $R(X) = 1$, where $R(X) = \wedge_{I \in \neg \Delta} \left( \sigma_I(X) \wedge F_I(X) \wedge \left( \wedge_{J \in LDOI(I)} \sigma_J(X) \right) \right)$ (and $R(X) \equiv 1$ when $\Delta$ is empty). This result is a necessary consistency condition for the inactivity of stable motif drivers. In the example of Figure 4, there is only one single-node driver of the system's sole stable motif, namely $(C, 1)$. Thus, the entire negated driver set is $\neg \Delta = (C, 0)$. We calculate that $LDOI(C, 0)$ is empty and that the update function for $(C, 0)$ is $F_{(C,0)}(X) = \neg X_A \vee \neg X_B$. Therefore, we find $R(X) = \neg X_C \wedge (\neg X_A \vee \neg X_B)$. In panel C of Figure 4, only the three yellow states have $R(X) = 1$, and so any motif-avoidant attractor is confined to those three states.

We also leverage time-reversal in determining terminality. If a stable motif $M^-$ of the time-reversal $TR(G)$ of the parity-expanded network $G$ is active in a state $X$, then $X$ can only be in an attractor of $G$ if $M^-$ (viewed as a set of virtual nodes in the forward-time system) is not self-negating in $G$. Thus, when considering states that may be in a motif-avoidant attractor of $G$, we can ignore states belonging to stable motifs of $G$, belonging to self-negating stable motifs of $TR(G)$, and states for which $R(X)$ is zero. We call the remaining states the *terminal restriction space* of $G$. For example, the terminal restriction space of the system in Figure 4 is the yellow portion of its state transition graph. In practice, focusing on the terminal restriction space can reduce the number of variables that must be simulated by a considerable amount. For example, we analyzed the 60-node T-LGL network of (*62*), whose STG is too large (hundreds of petabytes or more) to practically construct. Previous stable motif-based algorithms can estimate the attractor repertoire in a matter of hours, but cannot guarantee the absence of motif-avoidant attractors without building the STG (*48*). Exploring only the terminal restriction space, however, results in a reduction of the search-space by a factor of over two billion and allows us to exactly identify all the attractors of the system in a matter of minutes. (For the purposes of making explicit comparisons to a fully constructed STG, we analyze a simplified model of this same system, due to (*63*), in Text S6). We have also found many examples of random Boolean networks ($K = 2, p = 0.5, N = 4096$) in which the terminal restriction space of the initial network is of size zero; a state-space reduction of $2^{4096}$. For a description of the algorithm we use to efficiently search the reduced terminal restriction state space for motif-avoidant attractors, see Text S5.

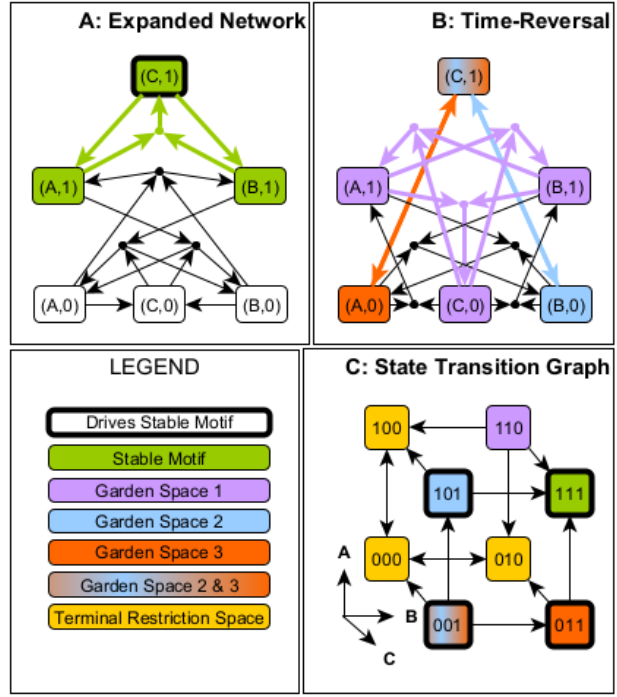

Figure 4: An illustration of methods to refine the portion of the STG that may contain motif-avoidant attractors using the system $f_A(X) = f_B(X) = \neg X_A \wedge \neg X_B \vee X_C; f_C(X) = X_A \wedge X_B$ as an example. Panel A presents the system's parity-expanded network and its sole stable motif (in green). This stable motif has one driver set, {(C,1)} (bold outline). By Corollaries 1 and 2 (Text S3), any motif-avoidant attractor must have $X_C = 0$ and $F_{(C,0)} = \neg X_A \vee \neg X_B = 1$ fixed, i.e., such an attractor can only contain states that satisfy $R(X) = \neg X_C \wedge (\neg X_A \vee \neg X_B) = 1$. Panel B shows the parity-expanded network of the time-reversed system, with its three stable motifs highlighted in purple, blue, and orange. Panel C shows the STG of the system. The states corresponding to each of the stable motifs in panels A and B are highlighted in matching colors. States with (C,1) active are highlighted with a bold outline; none of these states can be part of a stable motif-avoidant attractor. The three stable motifs of the time-reversal partition the STG into five subgraphs based on which time-reversal stable motifs are active: none, purple, blue only, orange only, or both orange and blue. These subgraphs, highlighted by color, are Garden of Eden spaces of the forward-time system. No attractor of the forward-time system can cross between these regions. Any motif-avoidant attractor of the system must reside in the terminal restriction space of the STG (yellow). Indeed, the states 100, 000, and 010 form a motif-avoidant attractor in which $X_C = 0$ is fixed while $X_A$ and $X_B$ oscillate.

In some cases, even the terminal restriction space is too large to simulate. In these cases, we can obtain information about the number of motif-avoidant attractors via the network reduction method of (*64, 65*). Variables that do not self-regulate are iteratively "deleted" by substituting their update functions into their successors' update functions (see Figure 5). This method, which we call *deletion projection*, provides a projection map, $\pi$, that has been proved to preserve certain features of the attractor repertoire, regardless of the number of deletions performed to obtain the projection map. In particular, all point attractors are preserved, and complex attractors of the original system map to one or more complex attractors of the projected system. Though not necessary for any of our theoretical results, variables with constant update functions are always prioritized for deletion as a computational optimization. It was shown in (*65*) that the order in which variables are deleted does not affect $\pi$. The inverse map preserves all point attractors, but only a subset of complex attractors in general (in other words, the projected system can have more, but never fewer, complex attractors than are present in the unprojected system). We combine the concepts of stable motifs and driver nodes with the deletion projection method of (*64*, *65*) to investigate the terminality of a motif-reduced Boolean system. We show that the domain of influence of a set of virtual nodes that specifies a state for every variable in the projected system has a domain of influence in the unprojected system that specifies exactly one state, called a "representative state" in (*64*) (Lemma 1, Text S3). Combining this result with Theorem 1 (Text S3) leads to one of our main results of this section: the activity of stable motifs in attractors is preserved by the deletion projection (Theorem 2, Text S3). Theorem 2 allows us to test the terminality of a system by testing the terminality of its projection.

We give a visual overview of the algorithm in Figure S2. We also present a "by hand" example of the approach on a five-variable system in Figure S3. We implement the techniques described in this section in the StableMotifs Python Library (available at https://github.com/jcrozum/StableMotifs/). Notably, our attractor-finding method outperforms the earlier stable motif-based method of (*48*) and the boolSim tool (*66*) integrated into GINsim (*67*) (see Text S8). For example, our code was able to compute the succession diagram of the 69-node unstimulated epithelial-to-mesenchymal transition model of (*33*) in under two minutes on a consumer-grade desktop, while the software implementation of (*48*) was unable to do so within twelve hours.

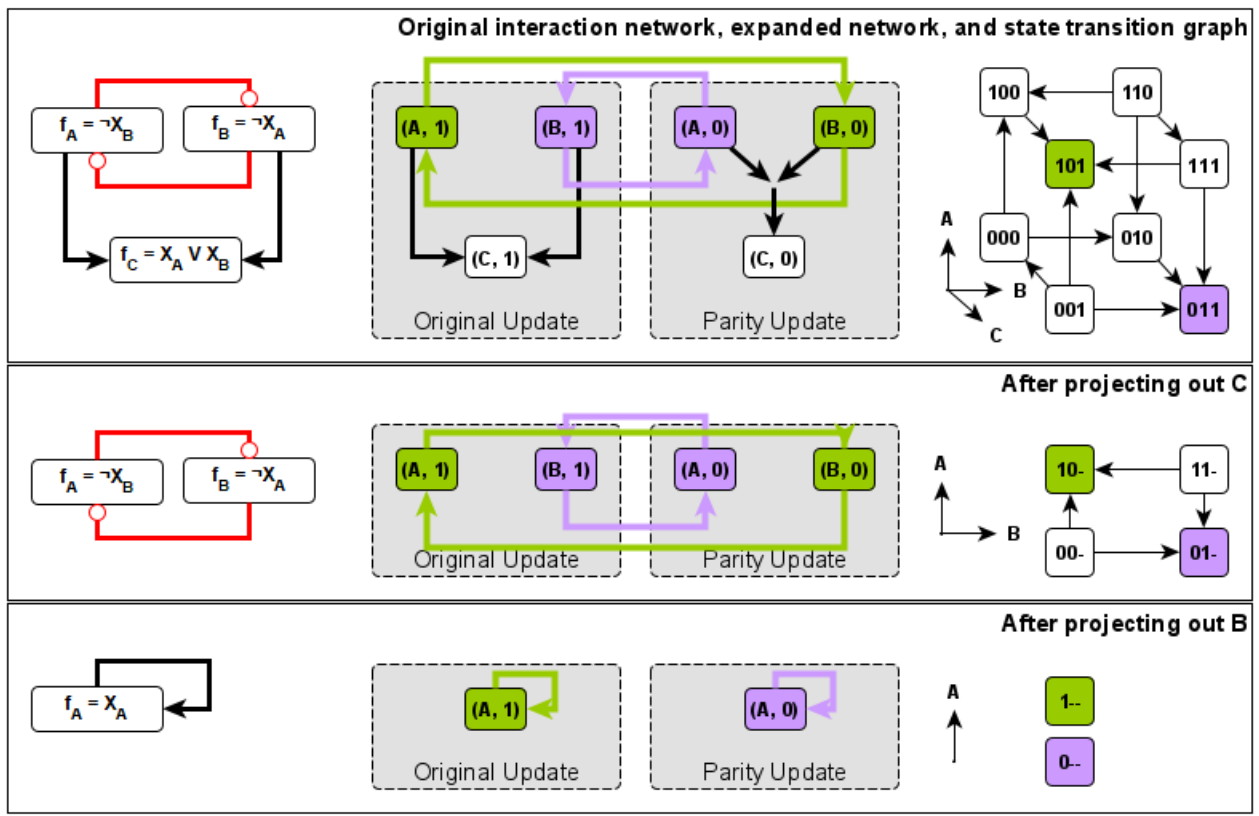

Figure 5: Deletion projection example. Each panel shows the interaction network, parity-expanded network, and STG of a system before projection (top), after projecting out one variable (middle), and after projecting out two variables (bottom). The attractors of each system and the stable motifs that are active therein are highlighted in green and purple. Attractors and stable motifs project onto attractors of the same color. The projection procedure preserves point attractors exactly. In general, the number of complex attractors (in this case zero) serves as an upper bound on the number of complex attractors in the unprojected system (see, e.g., (*64*)). One may view the action of the projection on the STG as contracting the top four nodes (100, 101, 110, and 111) to a single node (1--), with representative state 101 (transitions among these four nodes ultimately lead to 101); a similar view can be taken of the bottom four nodes. We note that the projection procedure respects the parity layer partitioning of virtual nodes. In accordance with Theorem 2, the stable motifs ({(A,1),(B,0)} and {(A,0),(B,1)}, which project to the self-activating virtual nodes (A,1) and (A,0), respectively) are preserved in the sense that, for example, the attractor with (A,1) active in the bottom panel corresponds to an attractor in which the stable motif {(A,1),(B,0)} (which projects onto (A,1)) is active. This implies that the value of $X_A$ is sufficient to determine which attractor the system attains.

## Results

### Application to decision-making in empirical biological network models

In this section we illustrate how our methods can be applied to validated Boolean models of biological networks to analytically connect regions of state space to decision-making (points-of-no-return) in their dynamics and to subnetworks (stable motifs) in the underlying interaction network. The biological network models we focus on are a model of the mammalian cell cycle Phase Switch (*68*) (presented in the subsection below) and a model (*63*) of a type of white blood cell cancer (T-LGL leukemia) (presented in Text S6). In particular, we identify and characterize the Garden of Eden spaces, illustrate an informative partitioning of the state space, and fully describe the attractor basins. The Phase Switch is a tri-stable molecular circuit that drives mitosis. Its three steady states mark three stages of the cell cycle: G1, G2, and the spindle assembly checkpoint (SAC). Under various biologically relevant conditions, such as coupling to other molecular circuits, the Phase Switch oscillates between these stages in the order they occur in the cell cycle (*50*) -- here, however, we study this switch in isolation. Further details of this model, including the update functions and stable motifs, are given in Text S7.

Our focus is on how details of a system's decision-making capacity can be explored by analyzing the system's time-reversal in conjunction with the succession diagram. We emphasize insights gained from the stable motifs of the time-reversed system. In principle, more granular initial condition tracing is possible via the full succession diagram of the time-reversed system. Figure 6 shows how the succession diagram and Garden of Eden spaces provide a concise summary of the irreversible commitments in state space; this is especially useful when the state transition graph is too large or too dense to analyze directly (compare panels B and E). The compressed state space illustration of Panel E unites information from the original and time-reversed system. It is equally or more informative than the full state transition graph (panel B). The green oval stands for Garden of Eden spaces, with the internal diamonds representing generic states in several categories of these spaces: the darkest green diamond represents Garden of Eden states (fixed points of the time-reversed system); the green diamond represents states in the overlap of multiple stable motifs in the time-reversed system; the light green diamond represents states in a single stable motif of the time-reversed system; and finally, the grey diamond represents states that do not lie in any stable motif of either the forward or backward time system. That states of this last type belong in a Garden of Eden space follows from the definition of stable motifs and the lack of motif-avoidant attractors in this system. Progression from darker to lighter green represents the commitment to exiting Garden of Eden spaces. Grey ovals represent overlapping spaces determined by the stable motif written inside each oval. The diamond symbols in these overlaps mark relevant trap spaces, e.g., the yellow diamond marks the trap space in which motifs P1, P5 and P6 are all active. Each edge corresponds to an irreversible commitment to a smaller trap space. For example, the yellow-blue diamond marking the intersection of the P1 and P5 spaces has two edges: the edge to the yellow diamond indicates a transition to the intersection of the P1, P5 and P6 motif regions, which determines an irreversible commitment to G2 (entry into the exclusive basin of attraction to G2), and the edge to the blue diamond indicates a transition to the intersection of the P0, P1 and P5 motif regions, which determines an irreversible commitment to G1 (entry into the exclusive basin of attraction to G1). Circular symbols indicate the attractors and highlight their position in the narrowest trap space. The graph structure in panel E is isomorphic to the graph structure in panel C, demonstrating that the stable motif succession diagram encapsulates the trajectory of the system in state space. The compressed succession-diagram representation has the benefit that it can be constructed for very large networks whose state transition graph cannot be built. Indeed, as we will demonstrate in the next section, we can build succession diagrams for networks whose state transition graphs have more nodes ($\approx 2^{4000}$) than can physically exist in the observable universe ($\approx 2^{600}$, assuming one state per Planck volume and disregarding graph edges).

An important feature of this analysis is that decisions can be ascribed to specific strongly connected subgraphs of the interaction network. These subgraphs and their stable states correspond exactly to stable motifs, i.e., their representations in the parity-expanded network do not have any parity-invariant subgraphs. These can provide important biological insights. For example, because there are no paths in the succession diagram from P1 to SAC (Figure 6C), we see that the P1 stable motif cannot be active during the spindle assembly checkpoint (SAC) phase. Driver node analysis of the Cdk1-Wee1 feedback associated to the P1 stable motif (Figure 6D) indicates that knockout of Cdk1 alone blocks the spindle assembly checkpoint. Following P1 commitment, the choice between G1 and G2 is decided by the activity of the Cdc25A-CyclinA feedback loop (P0) in competition with the Cdh1-CyclinA feedback loop (P6) when pAPC is inactive (P5).

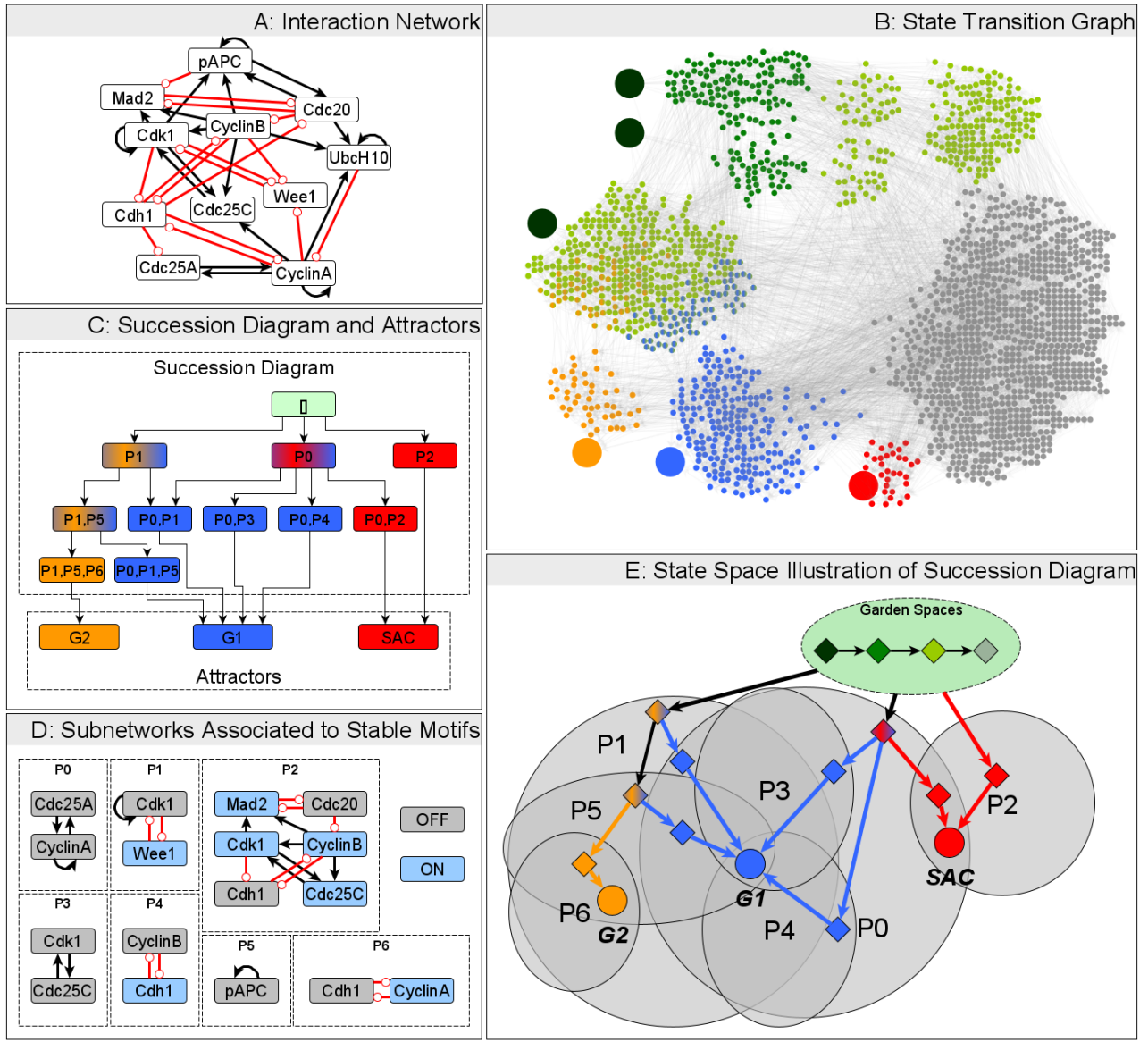

Figure 6: Parity and time-reversal analysis elucidates state-space decision-making in the Phase Switch model of (*68*). Text S7 provides further details about the model, including its update functions. The interaction network is presented in panel A. The full state transition graph is indicated in panel B. The three point attractors of this model are the three large nodes highlighted in blue, yellow, and red, and correspond to the G1, G2, and SAC phases of the cell cycle. The basins of attraction exclusive to each attractor are highlighted in the matching color. Grey nodes have paths to multiple attractors; the path taken depends upon the stochastic update order. The stable motifs of the time-reversed system identify unstable Garden of Eden regions of the state space (highlighted in green) which cannot be (re)entered from the outside. A darker shade of green indicates the overlap of multiple Garden of Eden spaces. The three large green nodes are the Garden of Eden states of the system, and are obtained analytically as the fixed points of the time-reversed system. Panel C shows the stable motif succession diagram of the system, with stable motifs of the network and reduced network denoted by labels P0-P6. The colors indicate, by the same scheme as in panel B, which attractors are possible after commitment to each stable motif. Panel D shows the subnetworks and node states associated with these stable motifs (blue: active/on, grey: inactive/off). These subnetworks and their states are obtained as strongly connected components of the parity-expanded network that do not contain parity-invariant subgraphs. In panel E, we illustrate how the succession diagram and garden spaces together describe (analytically) the possible decisions the system can make during its dynamics. Ovals represent specific subsets of the state space.

## Random Boolean network results

As an additional application we study the scaling of the average number of attractors of ensembles of RBNs generated by the Kauffman $N - K$ model (*24*). In this model, each of $N$ nodes receives $K$ edges from randomly selected "regulator" nodes. Each node's (quenched) update function randomly maps each of the $2^K$ possible combined regulator states to 1 with probability $p$ or to 0 with probability $1 - p$. Tuning the indegree $K$ or the activation bias $p$ yields an order-to-chaos transition at $2Kp(1 - p) = 1$ (in the thermodynamic limit, when $N \to \infty$) (*41, 69–71*). For additional details about previous studies of this model, see Text S1. We address and resolve the open problem of the attractor number scaling in the critical $K = 2$, $p = 0.5$ regime under stochastic asynchronous update. The number of attractors provably scales as a power law asymptotically bounded above by $N^{ln4}$ (*72*). In this section, we obtain the current best numerical estimate of the scaling exponent value (see Supplementary Code and the Random Boolean Network Application folder at https://github.com/jcrozum/StableMotifs/ for details on ensemble generation and analysis).

We generate ensembles for increasing network size $N$ (from size $N = 2$ to size $N = 4{,}096$) and apply our attractor identification algorithm to construct the succession diagram for each network and to determine or bound (in networks that are unusually computationally difficult) its number of attractors. For greater than 96% of the more than 10,000 unique networks generated we are able to exactly enumerate the attractors, and this fraction never falls below 86% (260/300 for $N = 2{,}048$) for any value of $N$ considered here. For all but 13 (0.1% of the total) of the networks without an exact attractor count we can constrain the number of attractors by using the deletion projection and/or counting the number of maximal stable modules (see Supplementary Code for details and implementation). Of these thirteen networks, the plurality (5) are of size 4,096, corresponding to 1.67% of the 300 networks generated for this size. For each of these remaining thirteen, we use the trivial lower bound of 1 for the number of attractors and choose an upper bound 10% larger than that of the most attractor-rich networks of the same size (see Supplementary Code for details). This makes these networks outliers without introducing undue sensitivity.

The scaling of the average number of attractors $\langle A \rangle$ as a function of network size $N$ is shown in Figure 7.

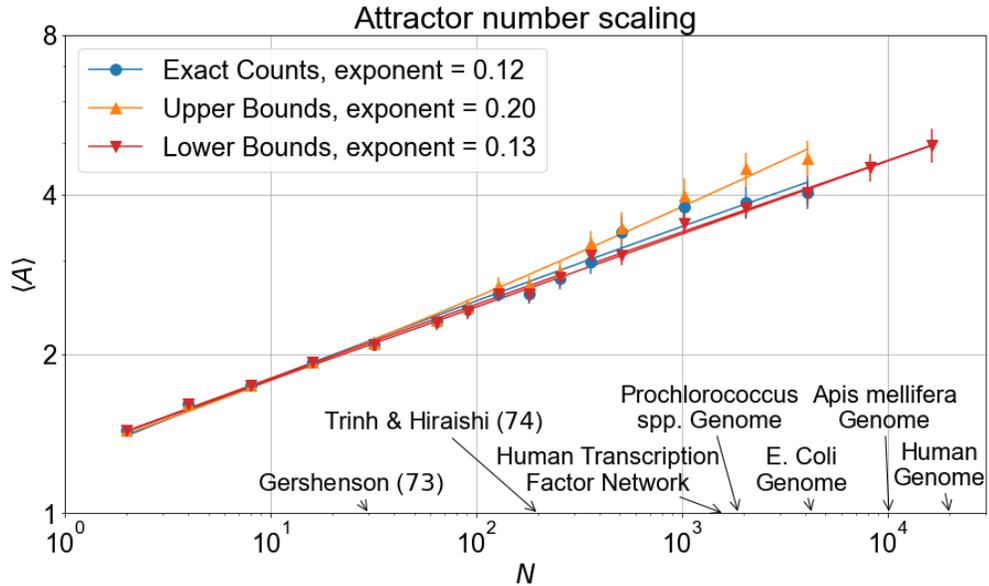

Figure 7: Summary of RBN attractor scaling fits. Symbols indicate the measured number of attractors and the lines represent fits of the form $\langle A \rangle = a + bN^c$ using nonlinear least-square fitting (see Supplementary Code for implementation details). The fits yield intercept $a = -0.38$, $a = 0.44$, and $a = -0.16$ for the exact counts, upper bounds, and lower bounds, respectively. The exponents of the fits ($c$ in $\langle A \rangle = a + bN^c$) are reported in the legend; see the main text or Supplementary Code for bootstrapped confidence intervals. For reference, the sizes of previously considered asynchronous RBN ensembles (*73, 74*) and several genetic networks from biology (*75–79*) are annotated on the horizontal axis.

By fitting power law $\langle A \rangle = a + bN^c$ we find that the exponent is $c = 0.12 \pm 0.05$ (one standard deviation; 95%CI [0.04,0.22]) when fitting only the networks for which we are able to exactly enumerate the attractors (blue circles and curve on Figure 7). The attractor upper bounds (orange symbols and curve) yield an upper bound on the scaling exponent of $c = 0.20 \pm 0.05$ (95%CI [0.11,0.30]). To ensure that the considered networks are sufficiently large, we analyze the scaling of the number of maximal stable modules in networks as large as $N = 16,384$ (recall that stable modules are source-free subgraphs of the parity-expanded network that have no parity-invariant subgraphs). Because maximal stable modules correspond to disjoint trap spaces in the state space, their number serves as a lower bound on the number of attractors. We include these lower bounds in our analysis of the lower bound scaling. In practice, the lower bounds are very often in 1-1 correspondence with the number of attractors, as is supported by the good agreement between the exact count of attractors and the lower bounds on the attractor counts. In particular, the scaling of the lower bounds ($c = 0.13 \pm 0.04$, 95%CI [0.06,0.21]) is consistent with the attractor scaling for $N \leq 4096$ and continues at least until $N = 16,384$, providing strong support that we have probed sufficiently large networks. These networks are of comparable size to many frequently studied genomes (e.g., $N \approx 4,000$ for the *E. coli* genome, while for the Human genome, $N \approx 20,000$). Furthermore, our analysis of these lower bounds increases our confidence that we have considered sufficiently large $N$ for the exact count scaling. This is because the mean estimates obtained by fitting up to $N = 4096$, $N = 8192$, and $N = 16,834$ are all in agreement with one another ($c = 0.13 \pm 0.04$ for each of them), and within less than 10% of the mean estimate obtained for the $N = 4096$ exact counts ($c = 0.12 \pm 0.05$). All the scaling exponents are well below the initially conjectured square root scaling ($c = 0.5$) and the theoretical maximum of $c = \ln 4 (\approx 1.39)$. For additional details regarding the fitting procedure and error estimation, see Supplementary Code.

Overall, we obtain the best current estimate of the exponent of $K = 2, p = 0.5$ Kauffman networks under stochastic update. Our analysis represents an 80-fold increase in network size over previous exact or near-exact enumeration analysis of asynchronous $N - K$ RBNs (*73,74*). We find a significantly lower exponent than the original exponent found by Kauffman and those identified in other stochastic updating schemes (*72-74*).

# Discussion

Two central questions in complexity science are "what emergent behaviors could a complex system exhibit?" and "what controls a complex system's selection of one emergent behavior or another?". While interesting questions from a purely theoretical perspective, they also have important applications in engineering, social science, and medicine. The past several decades have seen a growing number of collaborations between biologists, computer scientists, mathematicians, and physicists that approach these questions using Boolean networks. This work continues that interdisciplinary tradition, using geometric intuitions from physics to prove new results in the mathematics of Boolean dynamics, which we have applied to develop improved computational methods for analyzing complex systems common in biology and other sciences. Coming full circle, these methods have yielded new results in the statistical mechanics of random Boolean networks. Our methods allow efficient identification of the subspaces of the state space where robust commitments happen, and connect these decision-making spaces to subnetworks in the underlying interaction network. Time-reversal identifies a previously unexplored type of decision-making: the commitment to exit a Garden of Eden space, eliminating the option to ever return to that space. We associate each decision with the stabilization or de-stabilization of strongly connected subnetworks that do not contain any parity-invariant parts.

The parity symmetry of Boolean systems has led us to propose the parity-expanded network. Much like the state transition graph, it is parity-invariant (up to node relabeling) and completely encodes the system's dynamics, but its number of nodes grows linearly with the system dimension, rather than exponentially. It unites large knock-out and knock-in (constitutive activity) perturbations within the same framework, greatly simplifying their analysis compared to the similar ideas presented in (*31*), upon which we build. This earlier construction made use of three types of nodes and lacked explicit dynamical structure, requiring careful case-splitting and multiple system representations to study system perturbations (see e.g., *54*). Rather than distinguish nodes that correspond to on and off states at a definitional level, we distinguish these nodes by their behavior under a global parity transformation. This view elucidates the role of the parity-expanded network as a parity invariant and of its stable motifs as strongly connected components that have no parity-invariant subgraphs. These insights have enabled our formal proofs of novel results that relate network reduction methods, driver sets, and stable motifs -- results which we have leveraged to develop a fast attractor-finding method.

We have presented the time-reversal of a stochastic update Boolean system and demonstrated its usefulness in analyzing the forward-in-time dynamics. Our implementation is closely related to GINsim's model reversal (*67*). Just as stable motifs of a system describe stable spaces (subspaces that trajectories cannot exit) in the dynamics, stable motifs in the time-reversal of that system describe unstable spaces (subspaces that trajectories cannot enter). This observation is especially helpful in eliminating states when searching for attractors via direct STG construction, or in reducing the number of relevant initial conditions for study. In addition, it demonstrates an important property: the activity of any stable motif of a system or its time-reversal is constant for any attractor. In this way, time-reversal and parity elucidate the "attractor-conserved" quantities of a system's dynamics. These attractor-conserved quantities are only conserved within attractors and may initially vary. Nonetheless, it is surprising to note that in an inherently stochastic system there is a well-defined notion of time-reversal that yields asymptotic conservation laws.

By combining new results stemming from the parity-expanded network and time-reversal construction, we developed a method for fast attractor identification in stochastic update Boolean systems. We employed this new method to explore the scaling in attractor number for asynchronous $N-K$ Kauffman networks. Using these methods, we probed the power law scaling of the $K=2$ critical ($p=0.5$) Boolean networks under asynchronous update. Our new techniques allowed us to find or bound the number of attractors in these networks for sizes larger than ever before considered ($N=16{,}384$), and indeed we cover much of the biologically relevant ranges of gene regulatory network sizes. The power law scaling we observe is much lower (by a factor of about 10) than the theoretical maximum of $ln4$ (*72*), and also lower than the originally conjectured 0.5 scaling exponent (*24*). The low average number of attractors we find ( $\langle A \rangle \approx 4$ for networks with $N \approx 4{,}000$) is consistent with previous results on the average number of point attractors and stable synchronous attractors in this ensemble (*57*, *61*). Notably, the low attractor number and slow scaling appear even when upper bounds for attractor numbers are used in the calculations instead of exact attractor counts, meaning that the results are not explainable by a systematic undercounting of attractors in large networks. Extending the considered network size beyond $N=4096$ for the more easily computed lower bounds on the attractor size does not change the scaling exponent estimate. Because the exact counts and the lower bounds are in close agreement, this increases our confidence that we have considered sufficiently large $N$ for the exact counts of attractors.

The relatively slow growth of the average number of attractors compared to (i) the originally conjectured 0.5 scaling exponent and (ii) the current cell type scaling estimate of 0.70 (0.88/1.26 = 0.70, based on the experimental data from (*43*)) has several possible explanations that suggest follow-up investigations. One possibility is that timing-specific attractors contribute significantly to the cell-type scaling, implying that gene-regulatory synchrony plays a crucial role in a cell's ability to differentiate. Alternatively, the scaling of the attractor number under stochastic update might vary between critical RBN ensembles (e.g., it may differ in ensembles using canalizing functions or threshold functions), and some of these other ensembles could more accurately reproduce the observed cell type scaling. Analysis of the attractor number scaling in other RBN ensembles using our approach should help answer this question. It is also possible that gene regulatory networks of living organisms have evolved to increase the number of robust attractors, a process which is not fully captured by these ensembles of RBN.

The methods developed here can be readily applied to the numerous published Boolean models of biological systems to elucidate their full attractor repertoire. Our framework can also bring further insight into a variety of models that could be reformulated as Boolean models. For example, the quenched Glauber dynamics set on a network (*80*), models of binary opinion propagation (*8*), or the Hopfield model (*23*) can be expressed with threshold Boolean functions. The stable motifs of these systems, and correspondingly the trap spaces of their dynamics, can be identified as particular instances of strongly connected subgraphs. Indeed, in the Watts model of opinion propagation the percolation of an opinion depends on the existence of a strongly connected subgraph of early adopters, who can be influenced by a single neighbor to adopt the opinion (*18*). We expect that future adaptation of our methods to these models will be able to reveal rare attractors (metastable states).

Apart from the direct application to Boolean models we have emphasized here, time-reversal and parity play a role in describing the fundamental logical relationships between entities in complex systems more generally. In this view, the logical parity inversion and time-reversal of a system describe a coarse-grained and discretized version of the dynamics, which in turn provides insight into the dynamics of more detailed models; see e.g., (*45*, *81*) for further discussion. While the extent to which our key results generalize beyond the stochastic Boolean systems presented here remains an open question, we are encouraged by preexisting analogs of expanded networks in multi-level systems (*55*) and ODEs (*56*), as well as by results connecting logic-based models to ODEs (*45*, *81–83*). Though our focus here is at the level of interaction logic, our results suggest a new approach to analyzing complexity: studying the relationship of a complex system to its logical parity inversion and time-reversal to constrain the system's repertoire of emergent behaviors.

# Acknowledgements

**Funding:** This work was supported by NSF grants PHY 1545832, MCB 1715826, IIS 1814405 to R.A. and the Stand Up to Cancer Foundation/The V Foundation Convergence Scholar Award to J.G.T.Z. (D2015-039). **Contributions:** JR, JGTZ, and RA designed the research. JR and JGTZ developed the underlying theory with contributions from XG and RA. JR developed the software tools used with contributions from JGTZ and DD. JR, JGTZ, and DD conducted various numerical experiments. JR and JGTZ analyzed the simulation data with contributions from RA. JR and RA led the writing of the paper and preparation of the figures. **Data availability:** All data needed to evaluate the conclusions in the paper are present in the paper, the Supplementary Materials, and/or at https://github.com/jcrozum/StableMotifs. **Conflicts of interest:** All authors declare that they have no competing interests.

# Supplemental Material

Figure S1: The state transition graph of the example from Figure 3.
Figure S2: Flowchart illustrating the procedure for generating a succession diagram.
Figure S3: Application of the attractor identification algorithm on a five-variable example.
Text S1: Background on Random Boolean Networks
Text S2: Formal definitions of auxiliary networks
Text S3: Formal statements and proofs of key results
Text S4: Identifying terminal reduced networks using time-reversal and drivers
Text S5: Simulation and further reduction of the relevant state space
Text S6: Analysis of a T-LGL model using time-reversal
Text S7: Details of the cell cycle Phase Switch model of (*50*)
Text S8: Algorithm performance comparisons
Supplementary Code: Jupyter Notebook (Python 3) and input data for attractor scaling fits.
Supplementary Data: Benchmark raw timing data and plotting code.

# Supplementary Materials

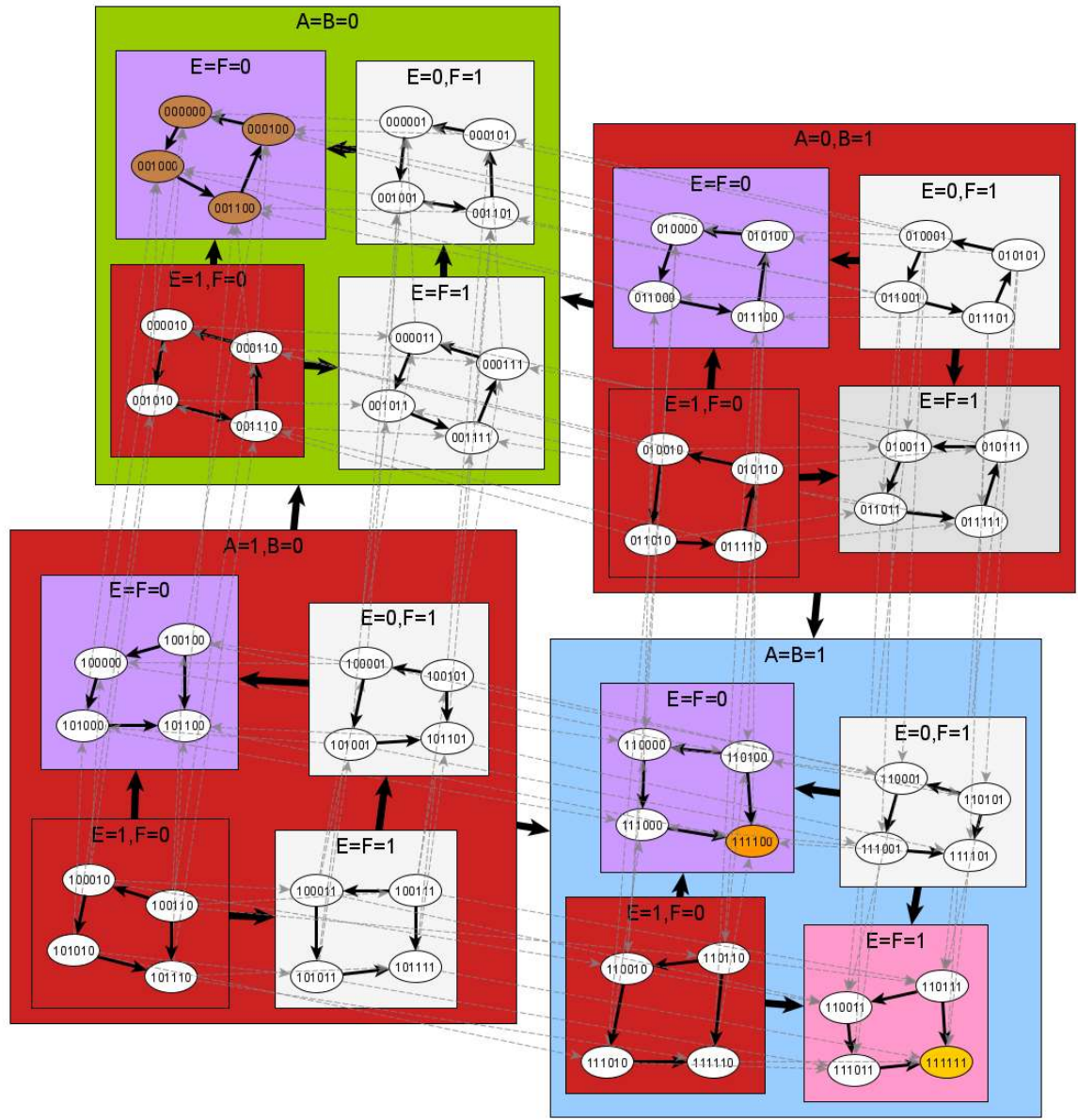

Figure S1: The state transition graph of the example from Figure 3. The state nodes are labeled according to the values of the variable $X_A$ through $X_F$, in alphabetical order. The states that make up the system's three attractors are highlighted in yellow, orange, and brown. State transitions are represented by grey dashed arrows, and the transitions between various groupings of nodes, or between nodes within a group, are summarized by solid black arrows. The states are grouped according to the nodes of the succession diagram in Figure 3E, with matching colors (green, purple, pink, and blue). Note that each of these state groups is a trap space in the sense that there are no transitions out of any of the stable motif regions. In addition, three additional groups are highlighted in red: the states in which $X_E = 0$ and $X_F = 1$; the states in which $X_A = 0$ and $X_B = 1$; and the states in which $X_A = 1$ and $X_B = 0$. These three groups

are the states in which the time-reversed system's three stable motifs are active. As such, these groups are Garden of Eden spaces. Indeed, there are no state transitions into any red group from outside that group. The state transition graph indicates that the system irreversibly leaves the Garden-of-Eden spaces and commits to increasingly restricted trap spaces. A particular example of a possible trajectory is to start from a state in the top right group of states (red box with A=0, B=1), enter the bottom right group of states (blue box with A=B=1), then enter the group of states with pink background (E=F=1), then converge into the attractor marked in yellow. This trajectory corresponds to the top branch of the stable motif succession diagram of Figure 3E.

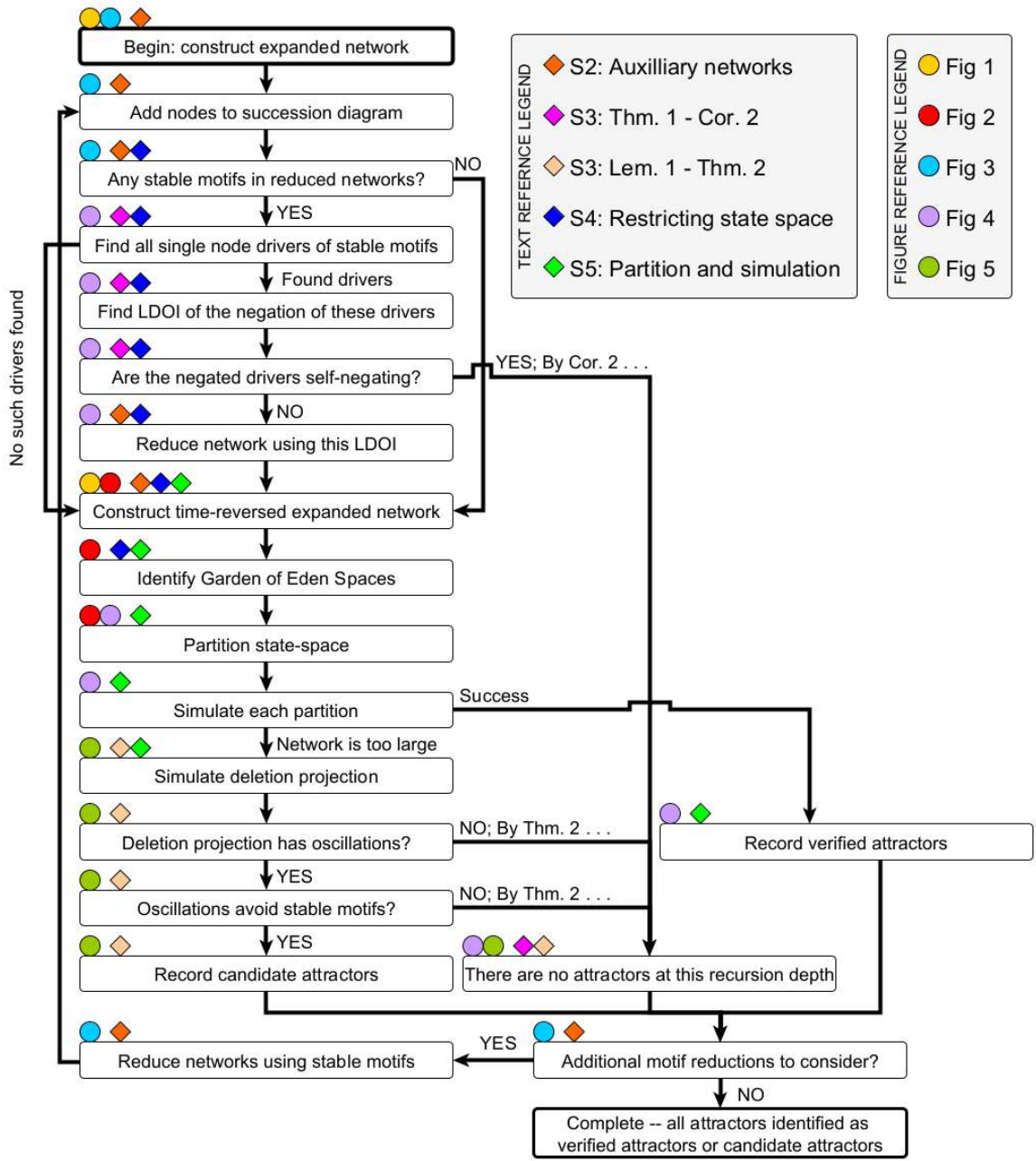

Figure S2: Flowchart illustrating the procedure for generating a succession diagram. Steps are tagged with colored circles indicating which figures provide additional detail for each step. The succession diagram construction proceeds as in Figure 3, but at each step in the recursion, we search for motif-avoidant attractors in order to identify the full attractor repertoire by using the methods summarized in Figures 4 and 5, which rely on special properties of driver nodes and Boolean time-reversal (Figure 2). In brief, the approach is to iteratively consider every possible sequence of stable motif activation (of any length, including zero). The system is simplified under the assumption that all stable motifs in the sequence are active. We then determine whether there are attractors that do not activate any stable motifs of the reduced system. If such an attractor exists, or if the reduced system does not contain stable motifs, the sequence of stable motifs is called terminal. The set of terminal sequences is in surjective correspondence with the attractors of the system.

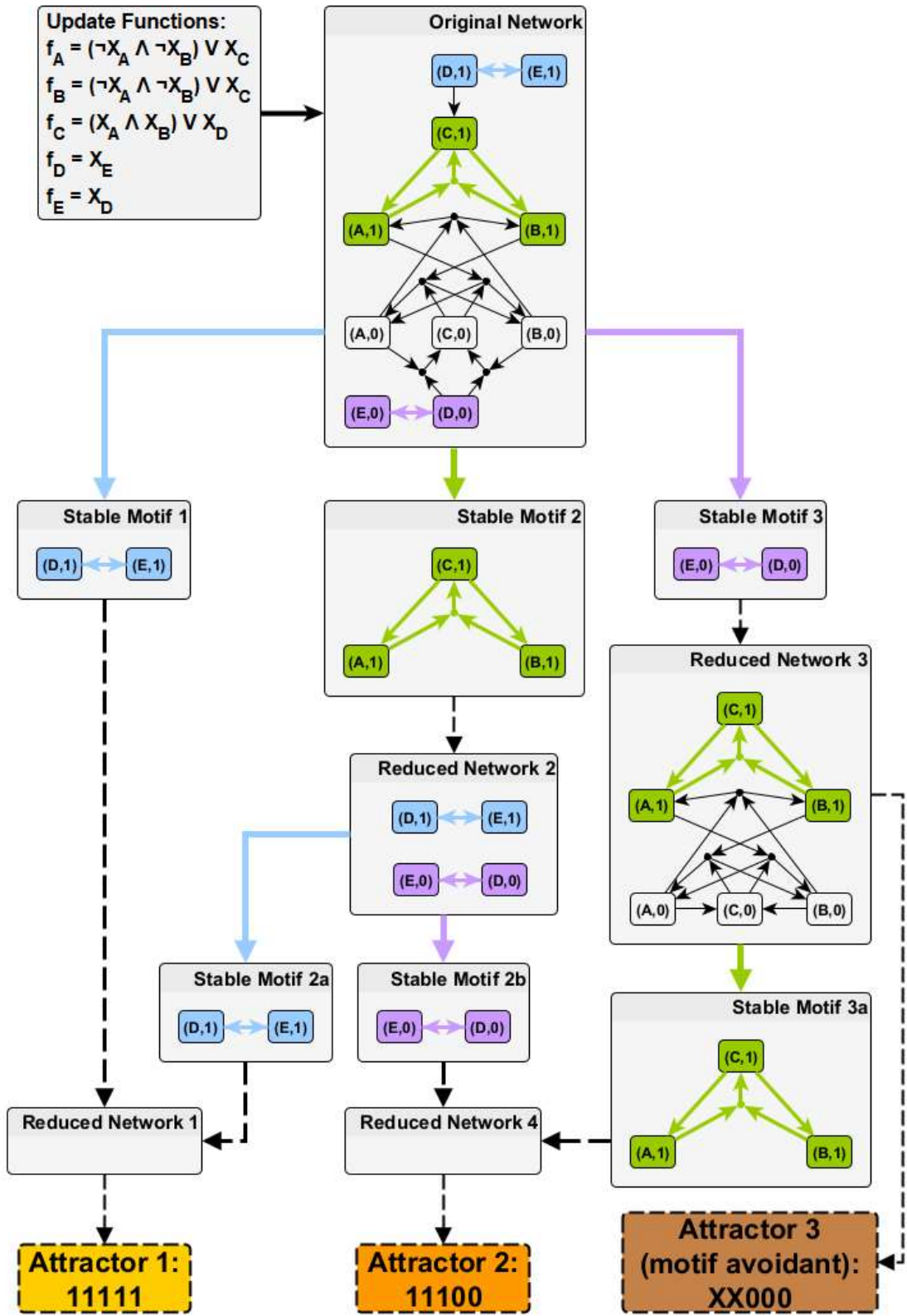

Figure S3: Application of the attractor identification algorithm on a five-variable example with update functions indicated in the top left panel. The parity-expanded network of the system is shown in the topmost center panel (Original Network), and below it are the three stable motifs it contains (labeled Stable Motif 1-3). Two of these arise from the E-D positive feedback loop and are mutually exclusive. Stable Motif 1 (blue) is formed by the virtual nodes $(D,1)$ and $(E,1)$, either of which can serve as a driver node of the motif. Stable Motif 3 (purple) is formed by the virtual nodes $(D,0)$ and $(E,0)$, either of which can serve as a driver node. Stable Motif 2 (green) is formed by the virtual nodes $(A,1)$, $(B,1)$, and $(C,1)$, the last of which is its driver node. The set of single-node drivers is thus $\Delta = (D,1),(E,1),(D,0),(E,0),(C,1)$, and so the negated single driver node set is $\neg\Delta = (D,1),(E,1),(D,0),(E,0),(C,0)$, which is self-negating by virtue of being self-contradictory. Therefore, at least one of the three stable motifs must lock in. After selecting a stable motif, the effects of its activation are depicted below it as a reduced network (labeled Reduced Network 1-3, respectively). This pattern is repeated iteratively until the resulting reduced network contains no further stable motifs, indicating that a minimal trap space has been uncovered. Each minimal trap space is typically small and therefore identifying the attractor(s) it contains is trivial. Attractors are listed along the bottom of the figure, with the binary strings indicating the states of the variables, in alphabetical order, that are fixed in each attractor; an 'X' in this string indicates oscillation. At each stage in the reduction process, one must consider whether there are motif-avoidant attractors that exist in a reduced network but fail to activate any of that network's stable motifs. Following the locking in of Stable Motif 1, the parity-expanded network reduces completely, and an "all-on" point attractor is obtained (Attractor 1); thus, there are no motif-avoidant attractors on this branch. If Stable Motif 2 activates in the original network, it yields the bistable switch shown in Reduced Network 2. This reduced network has two stable motifs (2a and 2b), either of which results in a fully reduced network and yields Attractor 1 or Attractor 2, respectively. Reduced Network 2 has negated single driver set $\neg\Delta = (D,1),(E,1),(D,0),(E,0)$, and thus, has no motif-avoidant attractors (i.e., either Stable Motif 2a or 2b must eventually activate). Finally, if instead Stable Motif 3 locks in, the example of Figure 4 is obtained (Reduced Network 3); this system has one motif-avoidant oscillation (Attractor 3, which involves the oscillation of two of the nodes in such a way that Stable Motif 3a never activates) and one point attractor (Attractor 2) that is reached if the Stable Motif 3a is activated rather than avoided.

## Text S1: Background on Random Boolean Networks

The Kauffman $N-K$ random Boolean network (RBN) model (*24*) has been frequently studied in systems biology and statistical physics for over 50 years. In this model, each of $N$ nodes receives $K$ edges from randomly selected nodes, and the (quenched) update functions are chosen such that each of $2^K$ input combinations yields an output of 1 with probability $p$. These models traditionally use the *synchronous update* scheme, in which all variables are updated simultaneously at each time-step, making the system's dynamics deterministic. Their attractors exhibit many features of biological cells, including stability against random external perturbations and plausible scaling laws for the number of attractors and attractor cycle lengths with the number of nodes $N$ (*43*, *84*). Specifically, tuning the indegree $K$ or the activation bias $p$ can produce an order-to-chaos transition at $2Kp(1-p) = 1$ (in the thermodynamic limit, when $N \to \infty$) (*41*, *69*). In the ordered regime, almost all nodes quickly attain a stationary state and small, transient perturbations tend to dissipate. In the chaotic regime, on the other hand, the number of fluctuating nodes is proportional to $N$ and perturbations grow exponentially fast. At $2Kp(1-p) = 1$ there is a so-called critical regime in which, on average, small perturbations persist indefinitely but remain small. RBNs with more heterogeneous topologies (*85*) or alternative distributions of update functions (*86–88*) also exhibit these regimes. The networks of greatest interest for biological systems were conjectured to lie near the critical regime (*84*). More recent work analyzing the dynamical behavior of gene regulatory networks (*89–91*) and studying evolutionary models of gene regulatory networks (*92*, *93*) has provided further support for this conjecture.

There has been a rich history of research on the scaling of the average number of attractors with network size in the biologically relevant critical regime; see (*43*) for a review of this work. Based on the relationship between the number of cell types and DNA content, Kauffman originally conjectured that the average number of attractors grows with the number of genes as a power law with exponent 1/2 (*24*, *84*). In the half-century since the introduction of the $N - K$ ensemble, numerical and theoretical analyses by multiple research groups have shown that the scaling of the number of attractors depends on the updating scheme and is not necessarily a power law (*70–72*, *94–97*). Under synchronous update, the average number of attractors for the $K = 2, p = 0.5$ critical Kauffman networks grows faster than any power law (in contrast to earlier conjectures) (*97*). Many of these attractors rely on perfect synchrony of multiple nodes. However, the assumption of synchronicity is not suitable for systems that exhibit multiple time scales or stochasticity (*57*, *98*). In contrast, under stochastic asynchronous update, the number of attractors grows as a power law, with an upper bound given by $N^{ln4} \approx N^{1.39}$ (*72*). The existence of a power law scaling in the asynchronous case is supported by a study of stable synchronous attractors (which are preserved under a perturbation of node update synchrony) in networks of up to $N = 40$ nodes (*57*). This study reported an exponent of 0.5 and a low average number of stable synchronous attractors (1.6 for $N = 30$). Another relevant result is that in $N - K$ ensembles with $p = 0.5$ the average number of point attractors is 1, regardless of the value of $N$ or $K$ (implying that for large $N$ at least a small number of nodes oscillate in most attractors) (*61*). Importantly, for the $K = 2$ critical Kauffman networks under stochastic asynchronous update, the exact value of the power law exponent is still unknown.

## Text S2: Formal definitions of auxiliary networks

The parity-expanded network $G$ is a dynamically endowed hypergraph constructed from a Boolean system's update rules and is defined as follows. Each node $I$ in the vertex set $V(G)$ of the parity-expanded network, called a *virtual node*, is an ordered pair $I = (n(I), s(I))$ consisting of a system entity, in this context denoted $n(I)$, and a value $s(I)$, which is either the constant 1 or the constant 0. In a Boolean model there are two virtual nodes associated with each system entity $i$, namely $(i, 1)$ and $(i, 0)$; we call this pair of virtual nodes contradictory. A set of virtual nodes that does not contain any contradictory pair is called *consistent*. We associate a variable $\sigma_I$ to each of the $2N$ virtual nodes. The function $\boldsymbol{F} = (F_{I_0}, \dots, F_{I_{2N-1}})$ maps $\boldsymbol{\sigma} = (\sigma_{I_0}, \dots, \sigma_{I_{2N-1}}) \in \{0,1\}^{2N}$ to an element of $\{0,1\}^{2N}$. In general, $\boldsymbol{F}$ is constructed for given update functions $f_i$ by writing each $f_i$ in Blake canonical form, then replacing all literals $X_j$ with $\sigma_{(j,1)}$ and all literals $\bar{X}_j$ with $\sigma_{(j,0)}$ to obtain $F_{(i,1)}$. The procedure is repeated with $\neg f_i$ to obtain $F_{(i,0)}$. In the example of Figure 1, $F_{(A,0)}(\boldsymbol{\sigma}) = \sigma_{(B,1)}$, $F_{(A,1)}(\boldsymbol{\sigma}) = \sigma_{(B,0)}$, $F_{(B,0)}(\boldsymbol{\sigma}) = \sigma_{(A,1)}$, $F_{(B,1)}(\boldsymbol{\sigma}) = \sigma_{(A,0)}$, $F_{(C,0)}(\boldsymbol{\sigma}) = \sigma_{(A,0)} \wedge \sigma_{(B,0)}$, and $F_{(C,1)}(\boldsymbol{\sigma}) = \sigma_{(A,1)} \vee \sigma_{(B,1)}$, where we may identify, for example, $F_{(A,0)}(\boldsymbol{\sigma}) = \sigma_{(B,1)}$ with $\neg f_A(\boldsymbol{X}) = X_B$ and $F_{(A,1)}(\boldsymbol{\sigma}) = \sigma_{(B,0)}$ with $f_A(\boldsymbol{X}) = \neg X_B$. Restricting the input so that $\sigma_{(j,0)} = \neg \sigma_{(j,1)}$ and requiring that $\sigma_{(j,0)}$ and $\sigma_{(j,1)}$ update together, using $F_{(i,0)}$ and $F_{(i,1)}$ respectively, restricts the dynamics to the $N$-dimensional subset where elements take the form $(\neg \boldsymbol{X}, \boldsymbol{X})$, in which case we write $F_I(\boldsymbol{X}) = F_I((\neg \boldsymbol{X}, \boldsymbol{X}))$ implicitly and define $\sigma_{(i,0)}(\boldsymbol{X}) = \neg X_i$, and $\sigma_{(i,1)}(\boldsymbol{X}) = X_i$, where we may think of $\sigma_I$ as the indicator function for the subspace defined by $I$. This is the context in which the evolution of by the action of $\boldsymbol{F}$ aligns with the underlying Boolean dynamics, and this is the context we consider throughout this work. For example, if we consider a virtual node $I = (i, 0)$, then $\sigma_I(\boldsymbol{X}) = 1$ holds when $X_i$ is 0 in state $\boldsymbol{X}$. A set of virtual nodes $S$ is *active* (in a system state $\boldsymbol{X}$) if all its members are active (i.e., if $\sigma_I(\boldsymbol{X}) = 1$ for all $I \in S$). The update function $F_I$ for each virtual node's activity is inherited from the update function for $n(I)$, i.e., $F_{(i,1)}(\boldsymbol{X}) = f_i(\boldsymbol{X})$ and $F_{(i,0)}(\boldsymbol{X}) = \neg f_i(\boldsymbol{X})$ are the update functions for $(i, 1)$ and $(i, 0)$; note that the activity of contradictory virtual nodes must always be updated together and have opposite states. A *hyperedge* connects a set of parent virtual nodes $S = \{I_0, I_1, \dots, I_k\}$ to a target virtual node $J$ if $\wedge_{I \in S} \sigma_I$ is a prime implicant of $F_J$.

The *succession diagram*, $\Sigma$, of a Boolean system is a directed acyclic graph whose nodes are the unions of the vertex sets of stable motifs used to obtain each reduced system. The reduced system corresponding to a succession diagram node exists in the trap space obtained by maximal percolation of the stable motif vertex sets that define the succession diagram node. Our software allows a choice to display or suppress the variable values that become fixed due to this percolation. For the sake of visual clarity, we do not display these values in the figures of this manuscript. Considering two nodes $P$ and $Q$ of the succession diagram $\Sigma$, there is an edge from $P$ to $Q$ if there is a stable motif $M$ in the reduction $Red(G,P)$ of an expanded network $G$ by a set of virtual nodes $P$ such that $Q = P \cup M$. For example, in Figure 3 of the main text the reduction by the blue stable motif (panel D) contains the pink stable motif, thus the succession diagram (panel E) contains an edge from the blue stable motif to the union of the blue and pink stable motifs. Each path in the succession diagram $p = (p_0 =, p_1, \ldots, p_n = P)$ from the empty set to a node $P \in \Sigma$ defines a stable motif history $M_i = p_{i+1} \setminus p_i$ for $i = 0..n-1$ and a corresponding sequence of reduced systems $Red(G, p_i)$. For each sink node of the succession diagram (i.e., whenever a reduced system has no additional stable motifs) there is guaranteed to be an attractor in which all reduced variables are fixed in their reduced states. If any variables remain unreduced, then one or more of the unreduced variables oscillate in the attractor. By repeating the reduction for every possible choice of stable motifs, one can construct a list of attractors of the Boolean system. For an example of this process, see Figure 3 of the main text.

## Text S3: Formal statements and proofs of key results

We begin by recalling the definition of the domain of influence and the closely related logical domain of influence.

*Definition: Domain of Influence (DOI).* We say that a set of consistent virtual nodes $S$ *drives* a virtual node $I$ if $I$ is consistent with $S$ and if $F_I(X) = 1$ for every attractor state $X$ of the dynamics obtained by restriction to the states in which $S$ is active. The set of all nodes driven by $S$ is called the *domain of influence* of $S$, and is denoted $DOI(S)$, or in the case that $S = \{(i,s)\}$, it can be denoted $DOI(i,s)$.

*Definition: multipath.* A (non-trivial) multipath from a set $S$ to a node $I$ is a (non-empty) finite sequence of hyperedges $\{(h_{parents,i}, h_{children,i}): i = 0,1,\ldots,n\}$ such that i) $h_{parents,0} \subseteq S = S_0$, ii) $h_{parents,i} \subseteq S_{i-1} \subseteq h_{children,i-1}$, and iii) $I \in h_{children,n}$.

*Definition: Logical Domain of Influence (LDOI).* A set $S$ of virtual nodes *logically drives* a virtual node $I$ if there is a non-trivial multipath in the parity-expanded network from a subset of $S$ to $I$ with all virtual nodes in the path consistent with $S$. The set of all $I$ logically driven by $S$ is called the *logical domain of influence* of $S$, written $LDOI(S)$ (or $LDOI(i,s)$ when $S = \{(i,s)\}$ is of size one). Intuitively, $LDOI(S)$ corresponds to the variable values that become fixed after percolating $S$ through the update functions and simplifying algebraically.

Note that it was shown in (*54*) that $LDOI(S) \subseteq DOI(S)$.

Theorem 1 and its corollaries relate an attractor to the domains of influence of driver sets that activate in said attractor.

**Theorem 1**: If an attractor $\mathcal{A}$ contains a state $X$ in which the set of virtual nodes $S$ is active, then $\mathcal{A}$ contains a state for which $DOI(S) \cup S$ is active.

*Proof of Theorem 1*: By definition, $\mathcal{A}$ contains all system states reachable from $X$ by all (stochastic asynchronous) update orders. In particular, it contains the set $\mathcal{A}'$ of all states reachable by update orders in which the variables described by $S$ are never updated and remain fixed at their values in $X$. Consider the modified system obtained by replacing the update functions $f_i$ for $X_i$ by the constant function $f_i(X) = s$ for all $(i,s) \in S$. By definition, the set $DOI(S)$ consists of all virtual nodes $(j,u) \notin S$ for which $X_j = u$ is fixed in all attractors of the modified system. It may also contain elements of $S$, but this is irrelevant when considering $\mathcal{A}'$ because $S$ is active in all states in $\mathcal{A}'$ by definition. In particular, $DOI(S) \cup S$ is active in all attractor states of the modified system. Because $S$ is active in both the modified system and along the paths from $X$ to the states in $\mathcal{A}'$, it follows that all paths that start from $X$ in the modified system can reach $\mathcal{A}'$. Therefore, $\mathcal{A}'$ contains an attractor of the modified system and thus a state for which $DOI(S) \cup S$ is active.

*Corollary 1*: If $DOI(i, s)$ contains a stable module $M$, then every attractor $\mathcal{A}$ in $STG(f)$ is either contained in the subspace in which $(i, \neg s)$ is active, or in the subspace in which $M$ is active (or both).

*Proof of Corollary 1*: If 2) does not hold in an attractor $\mathcal{A}$, then $\mathcal{A}$ contains a state $X$ in which $S = (i, s)$ is active. Thus, by Theorem 1, $\mathcal{A}$ contains a state in which $DOI(i, s)$, and $M$ in particular, activates. The permanency of the activation of stable modules (and motifs) implies that if $M$ is active in any state of $\mathcal{A}$, it is active in all states of $\mathcal{A}$. Therefore, 1) and 2) cannot both fail to be true.

The second corollary of Theorem 1 requires us to introduce the concept of *contradiction boundary*. If a virtual node $I$ is not consistent with a set of virtual nodes $S$, but there is a non-trivial multipath from a subset of $S$ to $I$ with all path nodes except $I$ consistent with $S$, then we say that $I$ is in the *contradiction boundary* of $S$. In this case the $LDOI(S) \cup S$ will contain a parent node of $I$, but it will not contain $I$. If $S$ is internally inconsistent (i.e., if it contains contradictory virtual nodes), then we say that the contradiction boundary of $S$ is the set of contradictory virtual node pairs in $S$. The contradiction boundary captures whether a set of virtual nodes $S$ will eventually activate a state that contradicts $S$. Recall that $S$ is self-negating if it drives a set $T$ that drives a node contradictory to $S$. Because $LDOI(S) \subseteq DOI(S)$, it follows that if the contradiction boundary of $S$ is non-empty, that $S$ is self-negating. The concept of contradiction boundary is important because of the following property:

*Remark 1:* If $S$ is self-negating, then every attractor contains at least one state in which $S$ is not active. In other words, no attractor lies entirely within the subspace defined by the activity of $S$.

*Corollary 2*: Let $\Delta$ be a set of virtual nodes such that for any $(i, s) \in \Delta$ there exists a stable module $M$ driven by $(i, s)$ i.e., with $V(M) \subseteq DOI(i, s)$. If there exists an attractor $\mathcal{A}$ in which none of these stable modules are active, then the image of $\Delta$ under parity, $\neg\Delta = \{(i, \neg s): (i, s) \in \Delta\}$, is active in every state of $\mathcal{A}$ and is not self-negating (and therefore its contradiction boundary is empty).

*Proof of Corollary 2*: Consider an arbitrary $(i, \neg s) \in \neg\Delta$. By assumption, $(i, s)$ drives some stable module (motif) $M$, and $\mathcal{A}$ lies outside the subspace defined by $M$. Because this holds for all $(i, \neg s) \in \neg\Delta$, it follows from Corollary 1 that the subspace defined by $\neg\Delta$ contains all of $\mathcal{A}$. Because no self-negating set of virtual nodes can be active in all states of any attractor, it follows that is not self-negating (Remark 1).

Lemma 1, below, relates the "representative states" represented in (*64*) to the DOI. More specifically, for update functions $f_i$ projected to a deletion projection $f_i^{red}$ by projection map $\pi$, each state $X_{red}$ of the projected system has a unique representative state in the unprojected system that all trajectories approach when confined to the subspace defined by $X_{red}$. The purpose of Lemma 1 is two-fold. First, we expand the unique representative state result in (*64*) to the case when multiple variables are deleted according to the algorithm of (*65*). The second purpose is to show that representative states can be interpreted as unique states specified by a particular DOI. This reinterpretation is useful in studying how stable motifs interact with the projection map.

*Lemma 1*: Let $G$ be the parity-expanded network of $f_i$, and $G_{red}$ be the parity-expanded network of a non-trivial deletion reduction $f_i^{red}$. If $S \subset V(G)$ transforms under parity to $V(G_{red}) - S$, then $DOI(S) \cup S$ specifies (i.e., is active in) exactly one state of $G$.

*Proof of Lemma 1*: Note that because $S$ transforms under parity to $V(G_{red}) - S$, it follows that $S$ is consistent and can be interpreted as a subset of $V(G_{red})$ that is active in exactly one state of $STG(f^{red})$, which we denote $X_{red}$. We construct a modified system $g_i$ as follows: let $g_i = f_i$ for $i$ corresponding to variables that are eliminated by projection, and $g_i = X_{red,i}$ for variables that remain after projection. Because $g_i$ and $f_i$ agree for all indices that are projected out by $\pi$, it follows that $\pi$ induces a deletion reduction of $g_i$ that results in $g_i^{red} = X_{red,i}$. Because $g_i^{red}$ is a constant function, it has exactly one attractor: the point attractor $\{X_{red}\}$. Thus, by the point attractor correspondence theorem of (*64,65*), $g_i$ has exactly one attractor, a point attractor $X$ that satisfies $\pi(X) = X_{red}$. Furthermore, because $S$ only constrains variables that remain after projection, $STG(g)$ and $STG(f)$ agree on the subspace defined by $S$. From the definition of domain of influence, this implies that $DOI(S) - S$ (the maximal subset of $V(G) - S$ that is active in all attractor states of $STG(f)$ restricted to the subspace defined by $S$) can be calculated using the parity-expanded network of $g_i$. Because $STG(g)$ restricted to the subspace defined by $S$ contains only the point attractor $X$, it follows that i) $DOI(S) - S$ is active in $X$ and ii) that if $j$ is the index of a deleted variable, there is exactly one virtual node $J \in DOI(S) - S$ with $n(J) = j$ (notably, this implies that $DOI(S)$ cannot be empty unless $\pi$ is the trivial identity map). Note also that every variable that is not deleted is similarly represented by exactly one virtual node in $S$ (which is also active in $X$). Therefore, every variable of $f_i$ has exactly one corresponding virtual node in $(DOI(S) - S) \cup S = DOI(S) \cup S$ and thus $DOI(S) \cup S$ is active in exactly one state: the point attractor state $X$ of $g_i$.

This lemma is closely related to various results in (*64*) regarding what are therein called "representative states". Lemma 1 tells us that fixing the state of the variables specified by S (which is given by the state of all variables of the projected system) will result in a point attractor in the unprojected system. It leads to a key result about the relationship between the deletion projection method and stable motifs: the activity of stable motifs in attractors of the projected system is indicative of their activity in attractors of the original system and vice versa. Theorem 2 makes this correspondence precise.

**Theorem 2**: Let $G$ be the parity-expanded network of $f_i$, and $G_{red}$ be the parity-expanded network of a non-trivial deletion reduction $f_i^{red}$ with projection map $\pi$. Let $M$ be a stable motif of $G$, and $\mathcal{A}$ be an attractor of $STG(f)$. Further, let $\mathcal{A}_{red}$ be an arbitrary attractor of $STG(f^{red})$ with $\mathcal{A}_{red} \subseteq \pi(\mathcal{A})$. The following implications hold:
1. If $M$ is active in any state of $\mathcal{A}$ then each state of $\mathcal{A}_{red}$ can be lifted to a state with $M$ active (i.e., $M$ is active in some state of $\pi^{-1}(X_{red})$ for every state $X_{red} \in \mathcal{A}_{red}$) and
2. If any state $X_{red}$ in $\mathcal{A}_{red}$ can be lifted to a state with $M$ active (i.e., if $M$ is active in some element of $\pi^{-1}(X_{red})$), then $M$ is active in every state of $\mathcal{A}$.

*Proof of Theorem 2*: First, note that $\mathcal{A}_{red}$ exists by Theorem 1, part 3 in [28]. Next, the two parts of the theorem are proved separately.
1. Assume $M$ is active in some state of $\mathcal{A}$. Then, because the activity of $M$ defines a trap space, $M$ is active in every state of $\mathcal{A}$. Consider any state $X_{red} \in \mathcal{A}_{red}$. Because $\mathcal{A}_{red} \subseteq \pi(\mathcal{A})$, there exists $X \in \mathcal{A}$ such that that $\pi(X) = X_{red}$. Because $M$ is active in every state of $\mathcal{A}$, it is in particular active in $X$. Therefore, $X$ is a state in $\pi^{-1}(X_{red})$ in which $M$ is active.
2. Assume $\mathcal{A}_{red}$ contains a state $X_{red}$ such that $M$ is active in some state $X \in \pi^{-1}(X_{red})$. Consider the set $S$ of $dim\, f_i^{red}$ virtual nodes in the parity-expanded network of $f_i$ that are active in all states that map to $X_{red}$ under $\pi$. By Lemma 1, $DOI(S) \cup S$ is active in exactly one state. Because $M$ and $S$ are both active in $X$ and $M$ describes a trap space, it follows that $DOI(S) \cup S$ cannot contradict $M$. Because $DOI(S) \cup S$ is active in only one state, this implies that $M$ is active in $DOI(S) \cup S$. Then from $X_{red} \in \mathcal{A}_{red} \subseteq \pi(\mathcal{A})$, it follows that $\mathcal{A}$ contains at least one state in $\pi^{-1}(X_{red})$, and thus it contains a state in which $S$ is active. Thus, by Theorem

1, $\mathcal{A}$ contains a state in which $DOI(S) \cup S$, and $M$ in particular, is active. Because $M$ describes a trap space, there are no state transitions to states in which $M$ is inactive. Therefore, $M$ is active in all of $\mathcal{A}$.

The next set of results, culminating in Theorem 3, concern the application of Corollary 2 and related results to the identification of attractors that do not exist in a trap space defined by a maximal stable module. The example of Figure 4 in the main text provides an example of such an attractor. In the context of the attractor identification algorithm discussed in this work, these attractors are those whose states do not activate any stable motifs in a reduced system with stable motifs. Specifically, an attractor $\mathcal{A}$ is called *motif-avoidant* in a parity-expanded network $G$ if i) $G$ has stable motifs and ii) none of the stable motifs of $G$ are active in every state of $\mathcal{A}$.

*Lemma 2:* If there exists an attractor $\mathcal{A}$ that is motif-avoidant in $G$, then $\neg\Delta$ (as defined in Corollary 2) is active in every state of $\mathcal{A}$ and is not self-negating.

Lemma 2 follows immediately from Corollary 2: if no stable motifs of $G$ are active in $\mathcal{A}$, then in particular the stable motifs with single-node drivers in are not active in $\mathcal{A}$; thus Corollary 2 applies directly.

*Lemma 3*: Let $S$ be any set $S$ of virtual nodes that is active in all states $X$ of an attractor $\mathcal{A}$. Then $\sigma_I(X) = F_I(X) = \bigwedge_{J \in LDOI(I)} \sigma_J(X) = 1$ is satisfied for all $I \in S$ and $X \in \mathcal{A}$.

*Proof of Lemma 3*: Because all of $S$ is active in every attractor state $X \in \mathcal{A}$, any virtual node $I \in S$ is active as well i.e., $\sigma_I(X) = 1$ holds. Because $X$ is an arbitrary state of $\mathcal{A}$, all reachable states must also satisfy $\sigma_I(X) = 1$ as well. Therefore, the update function governing $I$ must also yield 1 in the state $X$, i.e., $F_I(X) = 1$ must hold. Finally, because $I$ is active in all states $X \in \mathcal{A}$, $\mathcal{A}$ must lie within the trap space defined by $S$ and the percolation of the variable values it corresponds to. Algebraically, this is stated $\bigwedge_{J \in LDOI(I)} \sigma_J(X) = 1$.

*Theorem 3*: If an attractor $\mathcal{A}$ is motif-avoidant in $G$, then for every state $X \in \mathcal{A}$, the function $R(X) = \bigwedge_{I \in \neg\Delta} \left( \sigma_I(X) \wedge F_I(X) \wedge \left( \bigwedge_{J \in LDOI(I)} \sigma_J(X) \right) \right)$ takes the value 1.

*Proof of Theorem 3*: By Lemma 2 $\neg\Delta$ satisfies the conditions for S in Lemma 3. Thus, for every $I$ and attractor state $X \in \mathcal{A}$, these imply that $\sigma_I(X) = F_I(X) = \bigwedge_{J \in LDOI(I)} \sigma_J(X) = 1$ is satisfied. Taking the conjunction of these three constraints gives $\sigma_I(X) \wedge F_I(X) \wedge \left( \bigwedge_{J \in LDOI(I)} \sigma_J(X) \right) = 1$. Taking the conjunction of this equation over all values $I$ gives $R(X) = 1$.

Thus, Theorem 3 defines an algebraic condition that must be satisfied by any state belonging to a motif-avoidant attractor.

The last two results in this section formalize the Garden of Eden space property of stable motifs in the time-reversal parity-expanded network and demonstrate that the activity of such stable motifs is preserved within attractors of the forward-time dynamics.

*Lemma 4*: The activity of a stable motif $M$ in the parity-expanded network of the time-reversal $f^-$ of a Boolean system $f$ defines a Garden of Eden space in $STG(f)$.

*Proof of Lemma 4*: Because $M$ is a stable motif in the parity-expanded network of the reverse-dynamics, it defines a trap space in $STG(f^-)$. Thus, there are no state transitions in $STG(f^-)$ out of this space. Because $STG(f)$ can be obtained by reversing the edges of $STG(f^-)$, this implies that there are no state transitions in $STG(f)$ into the subspace defined by $M$. Thus, $M$ defines a Garden of Eden space.

*Theorem 4*: The activity of a stable motif $M$ in the parity-expanded network of the time-reversal $f^-$ of a Boolean system $f$ is constant within any attractor of $STG(f)$.

*Proof of Theorem 4*: By Lemma 4, the activity of $M$ defines a Garden of Eden space in $STG(f)$. Thus, there are no paths in $STG(f)$ from any state $X_{out}$ with $M$ inactive to a state $X_{in}$ with $M$ active. Thus, if $M$ is inactive in one state of an attractor, it is inactive in all states of that attractor. If $X_{in}$ belongs to an attractor, then transitions of the form $X_{in} \rightarrow X_{out}$ are not possible because such a transition would require a path from $X_{out}$ back to $X_{in}$ (as $X_{in}$ is in an attractor), in contradiction of Lemma 4. Therefore, whenever an attractor contains a state in which $M$ is active, it can only contain state transitions to other states in which $M$ is active, thereby preserving the activity of $M$. Thus, for any attractor $\mathcal{A}$, $M$ is either active in all states $X \in \mathcal{A}$, or is inactive in all states $X \in \mathcal{A}$.

## Text S4: Identifying terminal reduced networks using time-reversal and drivers

In general, testing for terminality requires analysis of the system's state transition graph. This approach, however, is computationally expensive; the STG can have as many as $2^N N$ edges (i.e., for $f_i(X) = \neg X_i$, $i = 1..N$). To avoid excess computation in special cases in which terminality can be determined without state-space simulation, we conduct a series of simple tests using necessary or sufficient conditions for terminality for every reduced system $G' = Red(G, P)$ with $P \in \Sigma$.

The first and simplest of these tests is to examine the out-degree of $P$. If $P$ has out-degree zero in the succession diagram $\Sigma$, then $G'$ does not contain stable motifs and is trivially terminal. A second test is supplied by Corollary 2 and Theorem 3 (see Text S3). We consider the set $\Delta$ of all virtual nodes that individually drive any stable motif of $G'$. By Corollary 2, all members of $\Delta$ must be inactive in any motif-avoidant attractor. For example, in Figure 4, the virtual node $(C,1)$ is a driver of the stable motif marked in green, so $X_C$ must be inactive in any motif-avoidant attractor. To identify any attractors of $G'$ in which all of $\Delta$ is inactive, we consider the negated set $\neg \Delta$ and its logical domain of influence, $LDOI(\neg \Delta)$. If $G'$ is terminal, then it has an attractor in which these sets are both active. Furthermore, if $\neg \Delta$ is stably active in some attractor, then the update values of the variables constrained by $\neg \Delta$ must also be fixed in the corresponding states. For example, in Figure 4, because $X_C = 0$ must hold for all states in any motif-avoidant attractor, $f_C$ must also be zero in that attractor. We summarize these various non-terminality conditions by the algebraic constraint that all motif-avoidant attractor states of $G'$ satisfy $R(X) = 1$, where $R(X) = \wedge_{I \in \neg \Delta} \left( \sigma_I(X) \wedge F_I(X) \wedge \left( \wedge_{J \in LDOI(I)} \sigma_J(X) \right) \right)$; this is shown rigorously in the proof of Theorem 3 in Text S3. In the example of Figure 4, there is only one single-node driver of the system's sole stable motif, namely $(C, 1)$. Thus, the entire negated driver set is $\neg \Delta = (C, 0)$. We calculate that $LDOI(C, 0)$ is empty and that the update function for $(C, 0)$ is $F_{(C,0)}(X) = \neg X_A \vee \neg X_B$. Therefore, we find $R(X) = \neg X_C \wedge (\neg X_A \vee \neg X_B)$. In panel C of Figure 4, only the three yellow states have $R(X) = 1$, and so any motif-avoidant attractor is confined to those three states

## Text S5: Simulation and further reduction of the relevant state space

If we are unable to rule out the possibility that the reduced system $G$ is terminal, we conduct a state-space search for attractors that do not activate any stable motifs in $G$. Because we are only concerned with a subset of the state-space, we are able to avoid simulating some of the states for an increase in computational speed. We achieve this by keeping track of a set of system states in the state transition graph of $G$, $STG(G)$, that cannot be part of any motif-avoidant attractor. We denote this set of system states $K$. Initially, applying Theorem 3, we take $K$ to be the terminal restriction space of $G$, i.e., $K$ is initially the set of states with $R(X) = 1$. We take each state of the state transition graph that is not in $K$, i.e., each state $X \in V(STG(G)) \setminus K$ and compute its successors using the Boolean update rules. If $X$ has a successor in $K$, then $X$ and all its ancestor states in $STG(G)$ (i.e., all states from which we have previously found a path to $X$) are added to $K$. Similarly, applying Theorem 4, $X$ and its ancestors are added to $K$ if there is a stable motif $M^-$ of $TR(G)$ active in $X$, but inactive in one of its successors. The subgraph $STG(G) \setminus K$ of

$STG(G)$ necessarily contains any attractors that are in $G$ but none of its network reductions. These attractors are identified as the terminal strongly connected components in $STG(G) \backslash K$.

There are two major computational benefits to this approach to simulating a reduced STG. First, each stable motif of a reduced system $G$ constrains at least one variable, so the size of the state space is reduced by at least a factor of two for every stable motif in $G$. Second, for a given STG, the task of finding attractors is equivalent to the task of finding terminal strongly connected components, which scales linearly with the number of edges in the STG (*99*). As such, reducing the size of the directed graph that must be searched for attractors generally results in drastically improved performance. The degree to which performance is increased depends upon the final size of the set $K$, which itself depends on the stable motif structure of $G$ and of its time-reversal.

In cases for which other simulation methods are computationally prohibitive, we resort to the deletion projection method and application of Theorem 2 in order to investigate terminality of a motif-reduced Boolean system. We find the attractors of the projected system and test whether each projection attractor is inconsistent (in the sense of Theorem 2) with all stable motifs. In other words, we test whether all stable motifs are inactive in all states of each attractor's preimage under the projection map. If none of the projection attractors are inconsistent, then the reduced system from which the projection was constructed is necessarily not terminal. Otherwise, the deletion reduction has motif-avoidant attractors. If the motif-reduced system has a motif-avoidant attractor, it necessarily corresponds to one or more of the motif-avoidant attractors of the deletion projection. In this way, the number of motif-avoidant attractors of the deletion projection places an upper bound on the number of motif-avoidant attractors of the motif-reduced system.

## Text S6: Analysis of a T-LGL model using time-reversal

To illustrate the new insights that the parity-expanded network framework and the time-reversal system can bring, we analyze a simplified network model (*63*) of a type of white blood cell cancer (T-LGL leukemia) shown in Figure S4. This model considers the relationships between five proteins (denoted "cer", "disc", "fas", "flip", and "s1p") and the process of cell death (apoptosis, denoted "apo"). The governing update functions are as follows:

$$f_{apo}(X) = X_{disc} \lor X_{apo}$$
$$f_{cer}(X) = \neg X_{apo} \land X_{fas} \land \neg X_{s1p}$$
$$f_{disc}(X) = (\neg X_{apo} \land X_{fas} \land \neg X_{flip}) \lor (\neg X_{apo} \land X_{cer})$$
$$f_{fas}(X) = \neg X_{apo} \land \neg X_{s1p}$$
$$f_{flip}(X) = \neg X_{apo} \land \neg X_{disc}$$
$$f_{s1p}(X) = \neg X_{apo} \land \neg X_{cer}$$

Previous analysis of the model indicated that it has two attractors, one corresponding to a T-LGL (cancerous) state and one corresponding to the normal behavior (apoptosis). We find that the system has three stable motifs, indicated in Figure S4A. Two of these are mutually consistent and describe trap spaces that contain the cancerous state, while the third contains the apoptotic state. The time-reversed system contains two stable motifs (Figure S4B), which together partition the state space into four regions (Figure S4C). The virtual node sets that define two of these regions (R1 and R2) are self-negating, meaning that these regions cannot contain attractors. The stable motif {(apo,1)} is active in all states of region R4. The LDOI of this stable motif specifies that fixing apo=1 drives the rest of the variables into their inactive states (see the edges originating from the blue virtual node in Figure S4A). Thus, region R4 contains the point attractor corresponding to apoptosis, which is the larger bold-outlined dark-blue node in Figure S4E. The stable motifs colored in red in Figure S4A are active in the subset of region R3 that corresponds to irreversible commitment to the T-LGL point attractor. The flow between the four regions can be readily determined (see thick dashed arrows in Figure S4E). The flow between regions indicates that all system decisions about which attractor to select occur in regions R1 and R3. Any trajectory in R1 may result in the initiation of the apoptosis process ($X_{apo} = 1$) and thus exit R1 via R2, which ultimately leads to the apoptotic attractor. Otherwise, trajectories eventually exit into region R3. In region R3, there are two possibilities: one of the TLG-L stable motifs activates, or the system escapes the region into R4 and the apoptotic attractor. This second scenario is possible because the restriction of the system to R3 (by setting $X_{apo} = 0$) gives rise to a so-called conditional stable motif (*50*) {(cer,1),(fas,1),(s1p,0)}, that is self-sustaining in R3, but ultimately drives (disc,1), which in turn drives (apo,1) and exit from R3 into R4, where it is no longer self-sustaining. The four states in which {(apo,0),(cer,1),(fas,1),(s1p,0)} is active (colored in light blue in Figure S4E) are therefore in the exclusive basin of attraction of the apoptosis attractor (i.e. the set of states from which every trajectory leads to the apoptosis attractor).Thus, by combining the inter-region flow with the irreversibility of stable motifs and the properties of driver sets, we recapitulate the previously obtained exclusive basins of the two attractors (in light blue and dark blue for the apoptosis attractor and in red for the cancerous attractor in Figure S4E). In summary, analysis of the parity-expanded network and of the time-reversed system allows an efficient and informative partitioning of the state space and identifies which of the partitions may and which may not contain an attractor.

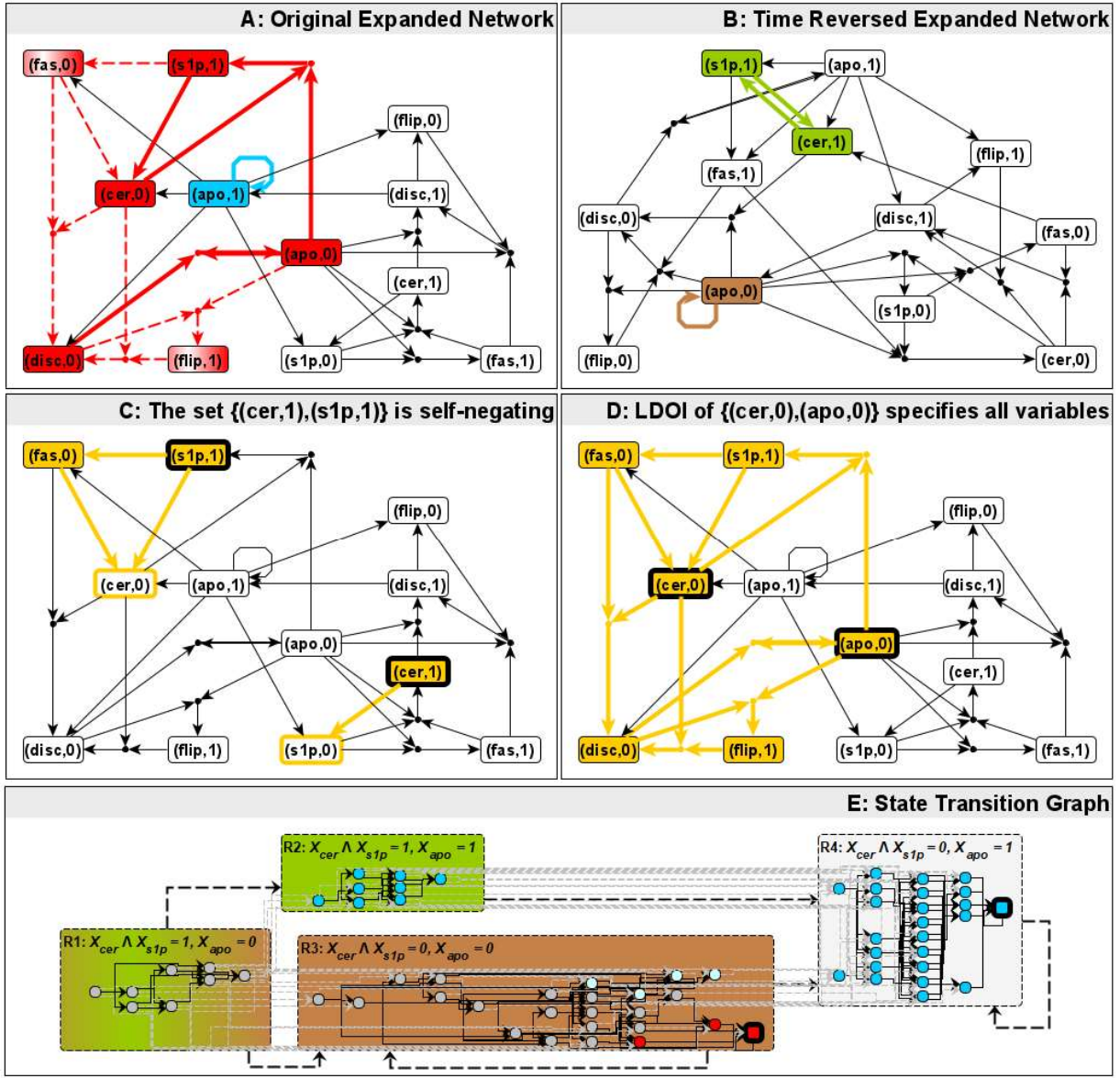

Figure S4: Time-reversal and domain of influence analysis of the simplified T-LGL leukemia network of (*63*). Panel A depicts the parity-expanded network of the system with its stable motifs highlighted in red and blue, respectively (the two partially shaded red nodes and the dotted hyperedges connected to them indicate that a stable motif is formed even when either (fas,0) or (flip,1) is omitted). Panel B depicts the system's time-reversal, with its stable motifs highlighted in green and brown, respectively. The stable motifs of the system's time-reversal allow us to partition the state-space into four regions according to which time-reversal stable motifs are active: R1) $X_{cer} \wedge X_{s1p} = 1$, $X_{apo} = 0$, R2) $X_{cer} \wedge X_{s1p} = 1$, $X_{apo} = 1$, R3) $X_{cer} \wedge X_{s1p} = 0$, $X_{apo} = 0$, and R4) $X_{cer} \wedge X_{s1p} = 0$, $X_{apo} = 1$. The properties of Garden of Eden spaces (which include R1, R1∪R2, and R1∪R3), imply that no attractor lies in more than one of these regions. Any attractor in R1 or R2 has {(cer,1),(s1p,1)} active, but this set of virtual nodes is self-negating. The LDOI of this set is depicted in yellow in panel C, with the white nodes with yellow border indicating the nodes that cause the set to be self-negating, i.e., these nodes are the set's contradiction boundary, as defined in Text S3. In any attractor in R3, either {(s1p,0),(apo,0)} or {(cer,0),(apo,0)} (or both) is active. The first of these is self-negating, but the LDOI of the second of these (depicted in yellow in panel D) fixes all variables, implying that R3 contains a single attractor: the point attractor corresponding to the highlighted virtual nodes. By a similar argument, R4, defined by the activity of either {(s1p,0),(apo,1)} or {(cer,0),(apo,1)} contains a single point attractor and no oscillations. The partitioning of the (forward time) STG according to the four regions is depicted in panel E, with each region colored according to which time-reversal stable motifs are active therein. States in the STG are colored in dark blue or red according to whether the stable motifs of the corresponding color are active in each state. The two point attractors are highlighted by bold outlines and are larger than other nodes. For the T-LGL attractor (red), the exclusive basin of attraction coincides with the activation of the corresponding stable motifs. In the apoptosis attractor (blue), the exclusive basin of attraction contains all states in which the apoptosis stable motif is active (dark blue), as well as four additional states (light blue) that can be identified by considering stable motifs of the system obtained by fixing $X_{apo} = 0$ (see text). Grey states may lead to either attractor depending on the stochastic update order. Solid black arrows indicate intra-region state transitions. Dashed arrows indicate inter-region transitions; the light grey arrows connect states in different regions, while the thick dashed lines summarize these state transitions as region-to-region transitions. A self-loop on a region indicates that it contains an attractor. Notably, the thick dashed lines illustrate the instability property of Garden of Eden spaces: once a trajectory exits the union of R1 and R2, for example, it can never return.

## Text S7: Details of the cell cycle Phase Switch model of (50)

The model of the cell cycle Phase Switch we consider here is an 11-variable network of interactions among proteins that govern mammalian cells' entry and exit from mitosis (cell division). The model was introduced in (68) and further analyzed in (50). The update functions are given below; we use the entity name and the name of its corresponding variable interchangeably (e.g., CyclinA should be understood to mean $X_{CyclinA}$ in the functions below).

$f_{Cdc25C}$= CyclinA or (CyclinB and Cdk1)
$f_{Cdh1}$= not CyclinA and (not CyclinB or not Cdk1)
$f_{Cdk1}$= Cdc25C and (CyclinA or CyclinB) and (Cdk1 or not Wee1)
$f_{CyclinA}$= (CyclinA or Cdc25A) and not (pAPC or Cdh1 and UbcH10)
$f_{CyclinB}$= (not pAPC and not Cdh1) or (not Cdc20 and not Cdh1)
$f_{pAPC}$= (pAPC and Cdc20) or (CyclinB and Cdk1)
$f_{Cdc20}$= pAPC and not Cdh1 and not Mad2
$f_{UbcH10}$= not Cdh1 or (UbcH10 and Cdc20) or (UbcH10 and CyclinA) or (UbcH10 and CyclinB)
$f_{Mad2}$= (not pAPC or not Cdc20) and CyclinB and Cdk1
$f_{Wee1}$= not CyclinA and not CyclinB or not Cdk1
$f_{Cdc25A}$=CyclinA and not Cdh1

This system has three point attractors, which correspond to distinct phases of the cell cycle, namely the G1 (first growth) phase, the G2 (second growth) phase, and the spindle assembly checkpoint (SAC). The G1 attractor is made up by Cdc25C = Cdk1 = CyclinA = CyclinB = pAPC = Cdc20 = UbcH10 = Mad2 = Cdc25A = 0, Cdh1 = Wee1 = 1. The G2 attractor is made up by Cdc25C = CyclinA = CyclinB = UbcH10 = Wee1 = Cdc25A = 1, Cdk1 = Cdh1 = pAPC = Cdc20 = Mad2 = 0. The SAC attractor is made up by Cdc25C = Cdk1 = CyclinB = pAPC = UbcH10 = 1, CyclinA = Cdh1 = Cdc20 = Wee1 = Cdc25A = 0.

The stable motifs of this system and of its reduced networks are labeled as follows
P0 := {(CyclinA, 0), (Cdc25A, 0)}
P1 := {(Cdk1, 0), (Wee1, 1}
P2 := {(Cdc20, 0), (CyclinB, 1), (Cdc25C, 1), (Cdk1, 1), (Cdh1, 0), (Mad2, 1)}
P3 := {(Cdc25, 0), (Cdk1, 0)}
P4 := {(CyclinB, 0), (Cdh1, 1)}
P5 := {(pAPC, 0)}

P6 := {(CyclinA, 1), (Cdh1, 0)}

Text S8: Algorithm performance comparisons

We compared the performance of the StableMotifs Python library, which implements the attractor algorithm presented in this work, to two other attractor identification codes: the earlier stable motif-based method of (*48*) and the boolSim tool (*66*). The algorithm of (*48*) was selected due to its similarity to the algorithm presented here and the algorithm of (*66*) was chosen because of its popularity; it is integrated into the widely-used GINsim (*67*) modeling software. All tests were conducted on a desktop computer with an Intel® Core™ i5-3330 CPU @ 3.00GHz × 4 with 7.8 GiB of RAM running Ubuntu 20.04.1 LTS, 64-bit. A main focus of this work is the analysis of Random Boolean Networks, and the critical $K = 2, p = 0.5$ networks in particular. As such, we use this ensemble for our primary point of comparison. We constructed twenty networks each of sizes $N = 10$ through $N = 100$ in increments of ten, for a total of two hundred networks. We identified the attractors in these networks using the three methods under consideration and recorded the time required for each network. We imposed a twelve-hour timeout, which was reached in several cases by both boolSim and the earlier stable motif-based software of (*48*), but never by the software presented here. The results are summarized in Figure S5. The software presented here performed significantly better than the other two tools for networks with more than $N = 50$ nodes, and in all cases the average time taken for our algorithm to identify the attractors remained below ten seconds.

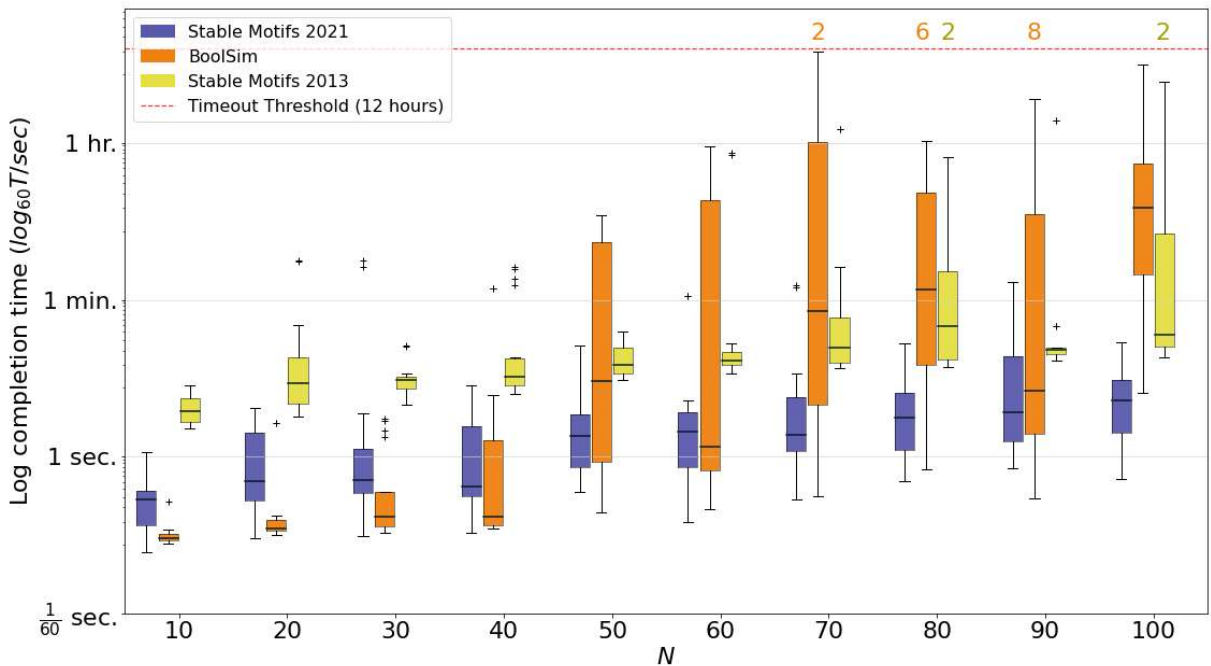

Figure S5: Comparison of attractor search times between different methods on an ensemble of random Boolean networks (RBN) of different sizes ($N$), ranging from 10 to 100 in steps of 10. For each N an ensemble of 20 different networks were created. The box and whiskers plots are generated from the base 60 logarithm of the time to analyze each network. Networks for which any of the three methods required greater than twelve hours to analyze are excluded from the statistics. In such cases, the number of networks exceeding the timeout threshold for each method is indicated above the corresponding distribution in a matching color. Notably, the method presented here never exceeded the twelve-hour threshold. The completion rates for the full ensemble of 200 networks are the following: Stable Motifs 2021: 100%; boolSim: 91%; Stable Motifs 2013: 98%. The raw timing data and plot generation code are available as supplementary material.

In addition, we compared the performance of the three software tools under consideration when identifying the attractors of three empirical networks. We analyzed the epithelial to mesenchymal transition (EMT) network of (*33*), the T-LGL survival network of (*62*), and the T-cell receptor signaling network of (*100*). The results are detailed below and summarized in table S1.

In the case of the 69-node EMT network, the earlier stable motif-based software of (*48*) was unable to complete the analysis in under twelve hours; boolSim identified all attractors in 5258 seconds (1.46 hours); the software presented here identified all attractors in 95 seconds. In the case of the 60-node T-LGL network, none of the software tools were able to identify all the attractors in under twelve hours. However, within 1481 seconds (i.e., just under 25 minutes), the method presented here was able to identify all 173 steady states and trap spaces containing at least 88 complex attractors in total. The 101-node T-cell signaling network of (*100*) was analyzed by the earlier stable motifs-based software of (*48*) in 4325 seconds (1 hour, 12 minutes), and by the newer method presented here in 45 seconds; 103 steady states and 25 complex attractors were identified. The attractor analysis using boolSim did not complete within twelve hours.

| Network | N | boolSim (*66*) | Stable Motifs 2013 (*48*) | Stable Motifs 2021 |
|---|---|---|---|---|
| EMT (*33*) | 69 | 5258 s | DNF | 95 s |
| T-LGL (*62*) | 60 | DNF | DNF | 1481 s (*) |
| T-Cell (*100*) | 101 | DNF | 4325 s | 45 s |

Table S1: Summary of attractor timing data for three empirical networks. Entries marked "DNF" indicate that the software did not complete analysis within the allotted twelve hours. The asterisk (*) in the middle entry of the rightmost column indicates the presence of trap spaces that may contain motif-avoidant attractors, but which were not explored.

## Supplementary Code: Jupyter Notebook (Python 3) and input data for attractor scaling fits.

## Supplementary Data: Benchmark raw timing data and plotting code.